\documentclass[twocolumn]{aastex631}
\usepackage{amsmath}
\usepackage{multirow}
\usepackage{tablefootnote}

\newcommand{\protag}{MAXI~J0655$-$013}
\newcommand{\code}{\texttt}
\newcommand{\comment}[1]{}

\accepted{July 12, 2023}

\submitjournal{ApJ}

\begin{document}

\title{Accretion spin-up and a strong magnetic field in the slow-spinning Be X-ray binary MAXI J0655-013}

\author[0000-0002-8403-0041]{Sean N. Pike}
\affiliation{Department of Astronomy and Astrophysics, University of California, San Diego, CA 92093}
\author[0000-0002-1190-0720]{Mutsumi Sugizaki}
\affiliation{National Astronomical Observatories, Chinese Academy of Sciences, 20A Datun Road, Beijing 100012, People's Republic of China}
\author[0000-0002-5686-0611]{Jakob van den Eijnden}
\affiliation{Department of Physics, University of Warwick, Coventry CV4 7AL, UK}
\author[0000-0003-0870-6465]{Benjamin Coughenour}
\affiliation{Space Sciences Laboratory, 7 Gauss Way, University of California, Berkeley, CA 94720-7450, USA}
\author[0000-0002-3850-6651]{Amruta D. Jaodand}
\affiliation{Cahill Center for Astronomy and Astrophysics, California Institute of Technology, 1200 E California Blvd., Pasadena, CA 91125}
\author[0000-0002-6337-7943]{Tatehiro Mihara}
\affiliation{High Energy Astrophysics Laboratory, RIKEN, 2-1, Hirosawa, Wako, Saitama 351-0198, Japan}
\author[0000-0002-6154-5843]{Sara E. Motta}
\affiliation{Istituto Nazionale di Astrofisica, Osservatorio Astronomico di Brera, via E. Bianchi 46, I-23807 Merate (LC), Italy}
\affiliation{Department of Physics, Astrophysics, University of Oxford, Denys Wilkinson Building, Keble Road, Oxford OX1 3RH, UK}
\author[0000-0003-0939-1178]{Hitoshi Negoro}
\affiliation{Department of Physics, Nihon University, 1-8-14 Kanda-Surugadai, Chiyoda-ku, Tokyo 101-8308, Japan}
\author[0000-0002-8808-520X]{Aarran W. Shaw}
\affiliation{Department of Physics, University of Nevada, Reno, NV 89557, USA}
\author[0000-0001-8195-6546]{Megumi Shidatsu}
\affiliation{Department of Physics, Ehime University, 2-5, Bunkyocho, Matsuyama, Ehime 790-8577, Japan}
\author[0000-0001-5506-9855]{John A. Tomsick}
\affiliation{Space Sciences Laboratory, 7 Gauss Way, University of California, Berkeley, CA 94720-7450, USA}



\begin{abstract}
We present MAXI and NuSTAR observations of the Be X-ray binary, \protag, in outburst. NuSTAR observed the source once early in the outburst, when spectral analysis yields a bolometric (0.1--100\,keV), unabsorbed source luminosity of $L_{\mathrm{bol}}=5.6\times10^{36}\mathrm{erg\,s^{-1}}$, and a second time 54 days later, by which time the luminosity dropped to $L_{\mathrm{bol}}=4\times10^{34}\,\mathrm{erg\,s^{-1}}$ after first undergoing a dramatic increase. Timing analysis of the NuSTAR data reveals a neutron star spin period of $1129.09\pm0.04$\,s during the first observation, which decreased to $1085\pm1$\,s by the time of the second observation, indicating spin-up due to accretion throughout the outburst. Furthermore, during the first NuSTAR observation, we observed quasiperiodic oscillations with centroid frequency $\nu_0=89\pm1$\,mHz, which exhibited a second harmonic feature. By combining the MAXI and NuSTAR data with pulse period measurements reported by Fermi/GBM, we are able to show that apparent flaring behavior in the MAXI light-curve is an artifact introduced by uneven sampling of the pulse profile, which has a large pulsed fraction. Finally, we estimate the magnetic field strength at the neutron star surface via three independent methods, invoking a tentative cyclotron resonance scattering feature at $44$\,keV, QPO production at the inner edge of the accretion disk, and spin-up via interaction of the neutron star magnetic field with accreting material. Each of these result in a significantly different value. We discuss the strengths and weaknesses of each method and infer that \protag\ is likely to have a high surface magnetic field strength, $B_{s}>10^{13}$\,G.
\end{abstract}



\section{Introduction} \label{sec:intro}
Each year, the Monitor of All-sky X-ray Image \citep[MAXI;][]{MAXI} observes dozens of astrophysical X-ray sources, often discovering previously unobserved sources when they enter periods of outburst. X-ray binaries (XRBs) represent a large portion of the sources observed by MAXI. XRBs are composed of a compact object --- a neutron star or a stellar mass black hole --- which accretes material from a companion star. As the accreted material spirals inward toward the compact object, gravitational potential energy is converted into radiative energy, reaching temperatures on the order of $10^7$\,K, making them strong X-ray emitters. The observable properties of an XRB, such as its spectrum and variability, depend upon the nature of the accretor, the parameters of the binary orbit, and the characteristics of the companion star.

On 2022 June 18, both MAXI/GSC (Gas Slit Camera) and the Neil Gehrels Swift Observatory's Burst Alert Telescope \citep[BAT;][]{Swift,BAT} detected the transient X-ray source, \protag, triggering the Nova-Alert System \citep{novaalert} and Hard X-ray Transient Monitor \citep{Krimm2013}, respectively. MAXI And BAT observations localized the source with an uncertainty of $\sim9^{\prime}$ and $3^{\prime}$, respectively \citep{Serino2022,Kennea2022b}. The flux and spectrum measured by MAXI suggested that the source was an XRB \citep{Serino2022}, and further monitoring with MAXI revealed apparent $\sim1$\,day periodicity as well as a power law spectrum with photon index $\Gamma=0.9\pm0.2$ \citep{Nakajima2022}. At the time of its detection, \protag\ was too close to the Sun for the majority of focusing X-ray observatories to perform the follow-up observations necessary to more precisely localize the source and to characterize its spectral and timing properties in detail. The Nuclear Spectroscopic Telescope Array \citep[NuSTAR;][]{NuSTAR}, however, is able to observe sources at a lower Sun avoidance angle than its peers. Therefore, NuSTAR performed follow-up observations of the source soon after its detection. These observations revealed strong, coherent pulsations with a period of $1129$\,s, indicating that the 1-day periodicity seen previously was likely due to sampling of this pulse period, and proving that the source consists of an accreting neutron star \citep{Shidatsu2022}. 

On 2022 August 20, when the source was no longer in close angular proximity to the Sun, Swift performed a follow-up observation, measuring a source count rate of $0.10 \pm 0.01\,\mathrm{s^{-1}}$ and precisely localizing the source within a 90\% confidence region centered on (J2000)  $\alpha=\mathrm{06h55m12.37s}$, $\delta = -01^{\circ}28^{\prime}52.7^{\prime\prime}$ with radius $2.5^{\prime\prime}$ \citep{Kennea2022a}. This localization demonstrated that the source is coincident with the previously catalogued quiescent X-ray source 2SXPS J065512.4$-$012855 \citep{Evans2020}, also known as SRGA J065513.5$–$012846, which has been associated with the 12.5\,mag variable star V520~Mon \citep{Pavlinsky2022}. A probabilistic analysis of the Gaia parallax reported by \citet{Vioque2020} combined with a direction-based prior places the system at a distance of $3.6^{+0.3}_{-0.2}$\,kpc \citep{BailerJones2022}. We adopt this distance throughout this paper. Optical observations of this companion star obtained in 2021 classified it as spectral type B1-3e III-V, indicating that \protag\ is an example of a Be X-ray Binary (BeXRB) \citep{Zaznobin2022}. Further optical observations in September, 2022, suggested that the companion has spectral type O9.5-B0V, confirming the classification of \protag\ as a BeXRB \citep{Reig2022}.

BeXRBs form a subcategory of high mass X-ray binaries (HMXB) primarily consisting of an accreting neutron star and a Be companion star. They are characterized by intervals of quiescence punctuated by outbursts during which their X-ray luminosities may approach or even exceed the Eddington limit. These outbursts can be separated into two categories \citep{Stella1986}. Type I outbursts occur at regular intervals corresponding to passage of the neutron star through periastron. These outbursts last several days and reach peak X-ray luminosities of $L_\mathrm{X}\lesssim10^{37}\,\mathrm{erg\,s^{-1}}$. Aperiodic type II ``giant" outbursts, on the other hand last much longer, sometimes longer than an orbital period, while reaching higher luminosities, $L_\mathrm{X}\gtrsim10^{37}\,\mathrm{erg\,s^{-1}}$. Giant outbursts may be caused by enhanced mass transfer due to eccentric or warped decretion disks around the Be star companions in highly misaligned BeXRBs \citep{Moritani2013,Martin2014}. 

Most BeXRBs exhibit X-ray pulsations, corresponding to neutron star rotation rates, with periods in the range of a few to thousands of seconds. Furthermore, they may exhibit cyclotron resonance scattering features (CRSF) in their X-ray spectra, providing a direct measurement of the magnetic field strength on the surface of the accreting neutron star. The rich spectral and timing features of BeXRBs make them an important laboratory for studying neutron star magnetic fields, transient accretion, and binary evolution. For a review of BeXRBs see \citet{Reig2011}.

In this paper, we present a detailed analysis of MAXI and NuSTAR observations of the BeXRB, \protag, in outburst. In Section \ref{sec:data} we describe the observations and data which we analyzed. In Section \ref{sec:outburst}, we present the long-term MAXI lightcurve, illustrating the structure of the outburst. In Section \ref{sec:timing} we present the timing features, including coherent pulsations, present in two focused observations taken with NuSTAR. In Section \ref{sec:spectra} we present a detailed spectral analysis of the NuSTAR observations of the source. In Section \ref{sec:pulseAverageLC}, we utilize the results of our NuSTAR analysis in combination with Fermi/GBM observations of the source in order to construct the pulse-phase-averaged MAXI light-curve. In Section \ref{sec:bfield}, we put constraints on the strength of the neutron star magnetic field using three different approaches, and in Section \ref{sec:conclusions} we discuss our results and summarize our conclusions. Throughout this paper, error bars are quoted at the 90\% confidence level unless otherwise stated.

\section{Observations and Data} \label{sec:data}

\begin{figure*}
\begin{center}
\includegraphics[width=0.40\textwidth]{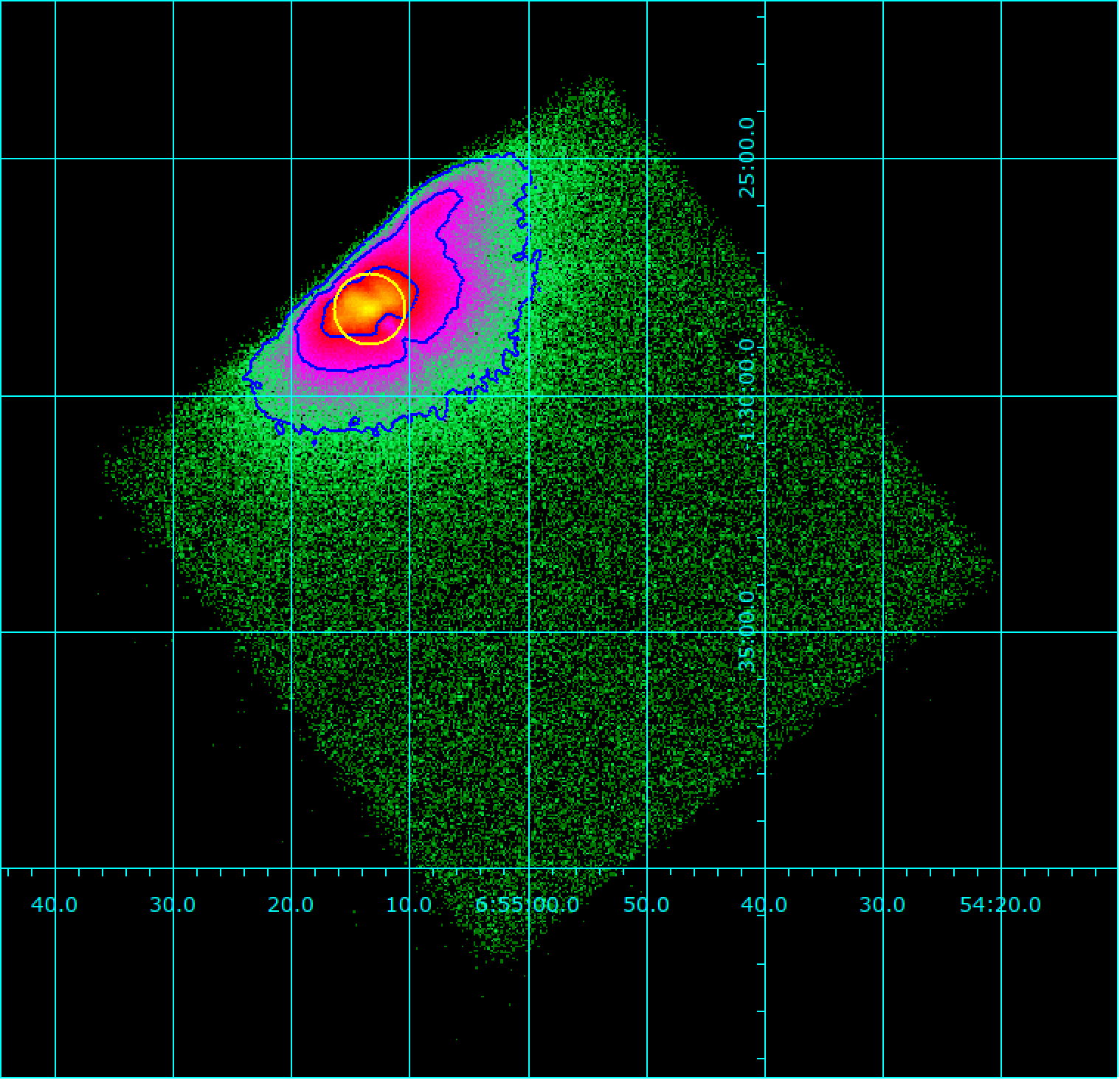}
\hspace{0.1\textwidth}
\includegraphics[width=0.41\textwidth]{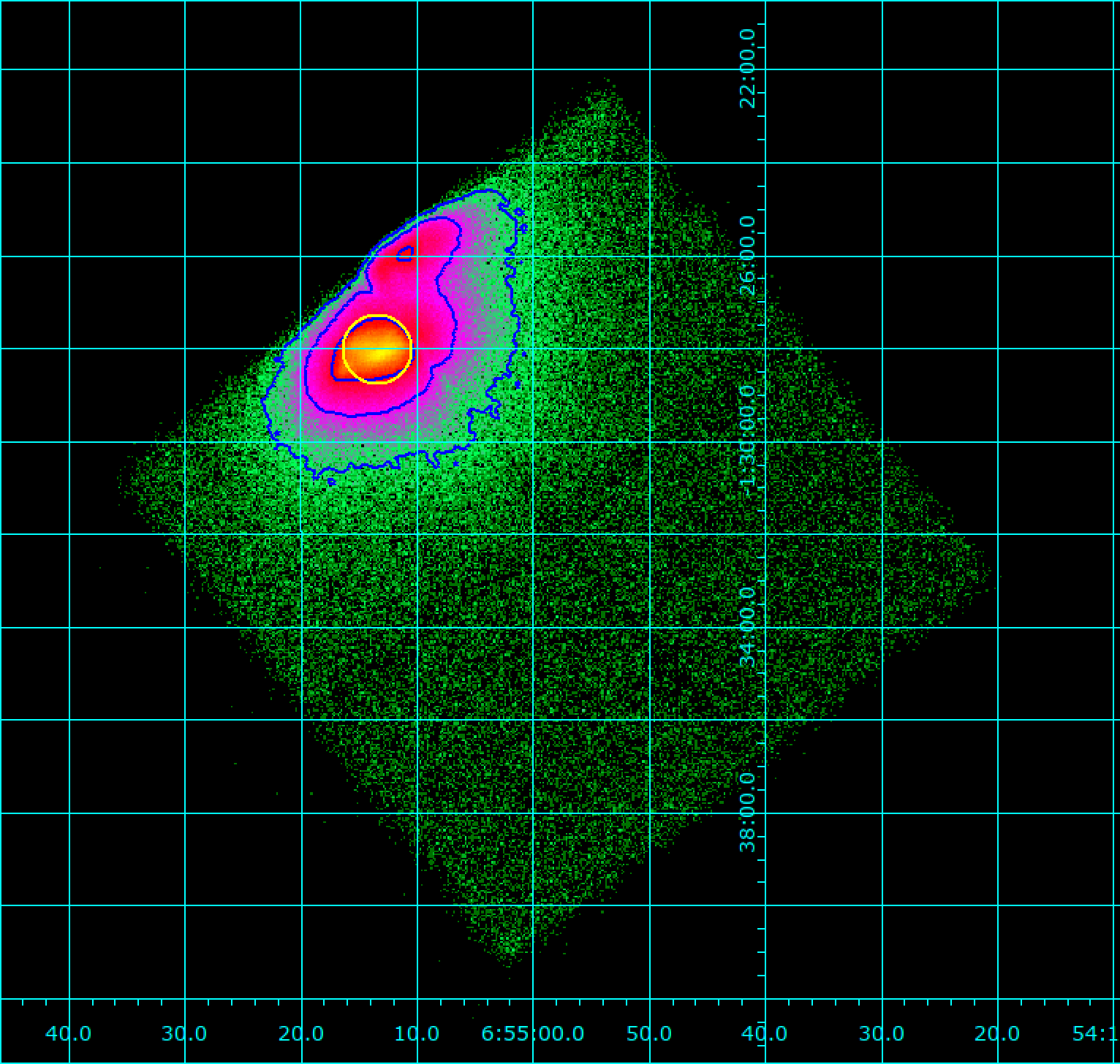}
\caption{Images reconstructed for FPMA (left) and FPMB (right) in Mode 6 for NuSTAR observation 80801347002. Note that due to poor astrometric reconstruction, source photons are spread into a region with multiple centroids. Additionally, during this observation the source fell near the edge of the detector and onto a gap between chips on FPMA. A circular region with radius $45^{\prime\prime}$ is shown in yellow for each FPM.}
\label{fig:Mode6_ims}
\end{center}
\end{figure*}

\begin{table}[]
    \centering
    \begin{tabular}{c|c|c}
        OBSID  & Start time (MJD) & Exposure time (ks) \\
        \hline
        80801347002 (OBS1) & 59751.94 & 44271 \\
        90801321001 (OBS2) & 59805.85 & 11335 \\
    \end{tabular}
    \caption{Details of the NuSTAR observations of \protag.}
    \label{tab:observations}
\end{table}

In this paper we present observations of \protag\ taken with MAXI and NuSTAR. Below, we describe the details of the observations and data extraction. For completeness, we also present the results of radio observations of the source.

\subsection{MAXI GSC}
MAXI is an all-sky X-ray monitor mission carried out on the
International Space Station (ISS). The main science instrument, Gas
Slit Camera \citep[GSC;][]{2011PASJ...63S.623M}, consists of gas
proportional counters and one-dimensional slat collimators. The
twelve identical camera units, named GSC\_0, ..., GSC\_9, GSC\_A,
GSC\_B, are arranged on the payload to cover two wide fields of view
of $160^\circ \times 3^\circ$, one facing at the earth horizon and one
at the zenith direction, simultaneously.
Since the instruments were activated on the ISS in 2009 August, MAXI
has been scanning almost the entire sky ($\gtrsim 80$\%) every ISS
orbital cycle ($\sim 92$ minutes). Here, we used the data of the seven GSC units which remain in normal operating condition.


We started data analysis with the GSC event data reduced from the data
transferred via the low-bit-rate downlink path in the 64-bit mode.
It covers the 2--20 keV energy band and 
has the best time precision of 50\,$\mu$s.
%
We utilized the standard analysis tools developed for the instrument
calibration \citep{2011PASJ...63S.635S}.
For each scan transit, the source event data were collected from a
rectangular region of $3\fdg 0$ in the scan direction and $3\fdg 6$ in
the anode-wire direction, with its centroid located at the position of
\protag. The backgrounds included in the source region were estimated
from the events in the same detector area, taken before and after the
source scan transits.
All event times recorded on the MAXI payload module 
were first converted to those at the solar system barycenter,
assuming all the photons came from the \protag\  direction and
referring to the ISS-orbit data.

\subsection{NuSTAR}
NuSTAR is the first hard X-ray observatory with focusing capabilities. It is composed of two optics modules with focal length 10\,m which focus X-rays onto two focal plane modules, FPMA and FPMB, each of which is made up of four cadmium zinc telluride (CZT) detectors. The resulting bandpass is $3-78$\,keV.

NuSTAR observed \protag\ twice. The first observation (OBSID 80801347002, PI Sean Pike; OBS1 hereafter) was taken beginning at UTC 2022 June 21, 22:33:36 (59751.94 MJD), while the second observation (OBSID 90801321001, PI Fiona Harrison; OBS2 hereafter) began at UTC 2022 August 14, 20:24:00 (59805.85 MJD). Details of the NuSTAR observations, including the OBSIDs, start times, and exposures times, are shown in Table \ref{tab:observations}. For both observations, we processed the data and extracted scientific products using the NuSTAR Data Analysis Software package (NuSTARDAS) v2.1.2 and CALDB v20220926. We produced cleaned and filtered event lists using \code{nupipeline} with no SAA or tentacle filtering, and we use \code{nuproducts} to produce science-level products such as light-curves and spectra.

OBS1 was taken in Mode 6 due to the proximity of \protag\ to the Sun at the time. Under normal operating conditions, known as Mode 1, NuSTAR uses two star-trackers, or Camera Head Units (CHU), attached to the optics modules in order to perform image reconstruction. However, when observing sources near the Sun, these two CHUs are unavailable. Therefore, we must rely on additional CHUs attached to the spacecraft bus in order to perform image reconstruction. This mode of image reconstruction is known as Mode 6. As the spacecraft moves, different combinations of these CHUs are functional. Due to complicated effects such as thermal flexing of the spacecraft bus, image reconstruction in Mode 6 is degraded compared to Mode 1, and the source may appear to ``move" around the focal plane depending upon CHU combination, resulting in multiple apparent centroids. In Figure \ref{fig:Mode6_ims} we show the images obtained during this observation, which demonstrate the effects of Mode 6 image reconstruction. During this observation, due to the fact that the source had not yet been precisely localized, \protag\ fell near the edge of the NuSTAR field of view, and near a gap between detectors. This is more pronounced for FPMA than for FPMB. In order to mitigate some of these effects when producing spectra and light-curves, we used the command \code{nusplitsc} in order to split the cleaned event files according to CHU combination. We processed each of these event files separately, choosing appropriate circular source and background regions for each, then we combined the resulting scientific products at the end. Essentially, for the purposes of light-curve and spectrum extraction with \code{nuproducts}, we treated each CHU combination as a separate observation. Because OBS2 was performed in the usual Science mode (Mode 1), we did not need to split the observation into different CHU combinations.

We utilized the software package DS9 \citep{ds9} in order to produce images and to select source and background regions. We used Stingray \citep{stingray_doi,Huppenkothen2019a,Huppenkothen2019b} to perform timing analyses, such as producing power spectra and searching for pulsations. We performed spectral modeling using Xspec \citep[v12.13.0c;][]{Xspec}.

\subsection{MeerKAT}

As part of the ThunderKAT Large Survey Program \citep{Fender2016b}, which routinely observes active Southern X-ray binaries, cataclysmic variables, supernovae, and gamma-ray bursts, we observed the position of \protag\ with the MeerKAT radio telescope. 
We visited the target over the course of two months, observing the field over 7 epochs on 2022 June 21 and 26, July 3, 9 and 17, and August 1 and 13. We used the telescope’s L band receivers, and we obtained data at a central frequency of 1.28 GHz across a 0.86 GHz bandwidth (856 – 1712 MHz). All the observations consisted of 15~min of on-source time, book-ended by two 2~min scans of the secondary calibrator (J0725-0054), plus a 10~min observation of a primary  calibrator (J0408-6545).

We conducted the subsequent analysis via a set of \textsc{Python} scripts specifically tailored for the semi-automatic processing of MeerKAT data (\textsc{OxKAT}\footnote{\url{https://github.com/IanHeywood/oxkat}}, \citealt{Heywood2020}). 
\protag\ was not detected in any of the images extracted from the individual observations. Therefore, in order to increase the image sensitivity and S/N, we stacked the available data and extracted one single image, which features a remarkably low rms noise in the vicinity of the target of 9.6$\mu$Jy. The target was not detected in the stacked image, and hence we can place a $3$-$\sigma$ upper limit to its radio flux to 29$\mu$Jy. At the distance of \protag, this corresponds to a 6\,GHz luminosity limit of $3.5\times10^{27}\,\mathrm{erg\,s^{-1}}$. Compared to radio monitoring of other BeXRBs in a similar X-ray luminosity range, no radio emission would be expected above this limit \citep{vdEijnden2022}.

\section{The shape of the outburst} \label{sec:outburst}

\begin{figure*}[t]
\begin{center}
\includegraphics[width=0.7\textwidth]{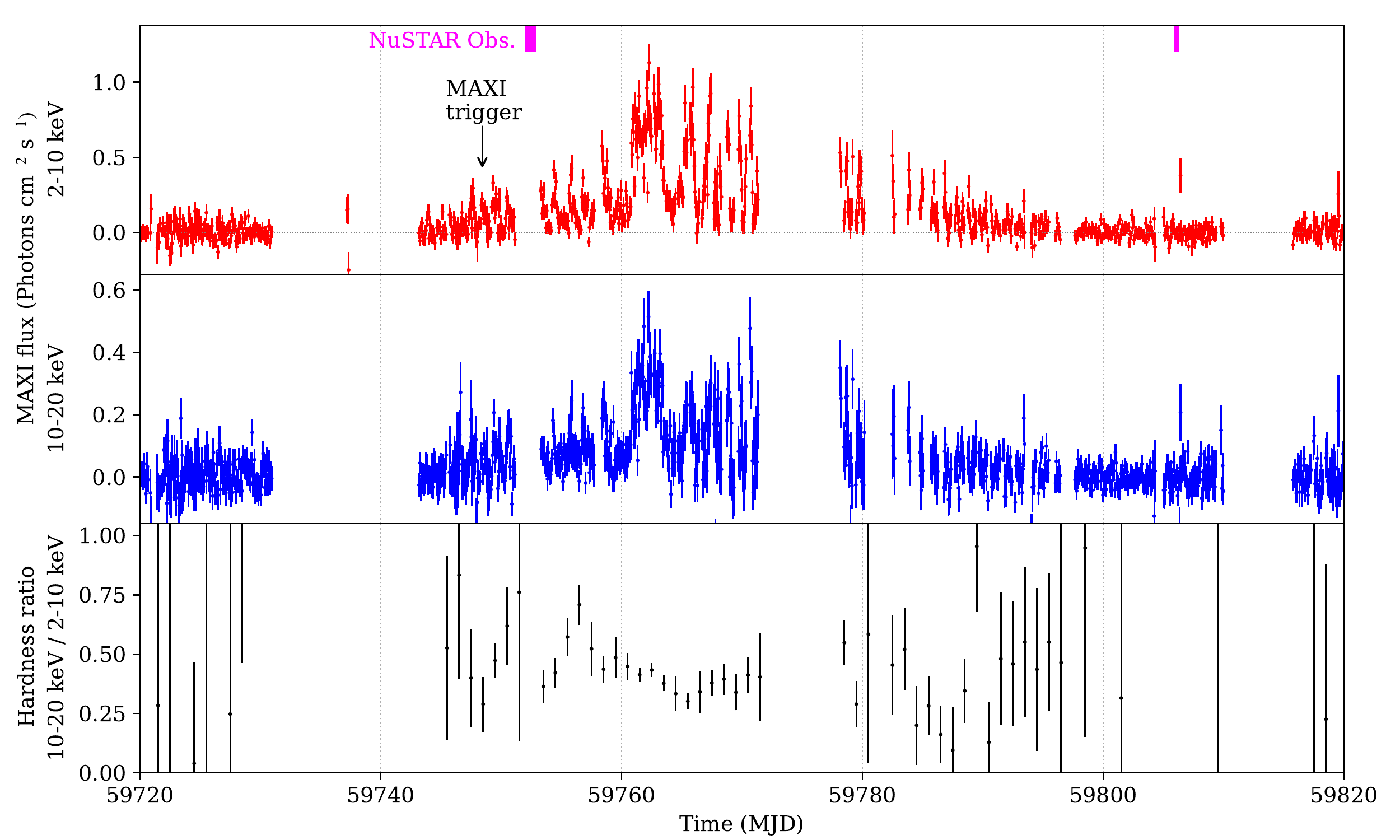}
\caption{\protag\ X-ray light curves for 100 days from 2022 May 21
  (MJD 59720) to August 29 (MJD 59820) in the 2-10 keV (top panel) and
  10-20 keV (middle panel) bands obtained by the MAXI/GSC with scan
  observations of $\sim 40$ seconds every ISS orbital cycle ($\sim 92$
  min). The hardness ratio of 10-20 keV flux to 2-10 keV flux in 1-day time
  bins is shown in the bottom panel. The intervals in magenta at the top panel
  indicate the two NuSTAR observations.
  %
}
\label{fig:maxilchr}
\end{center}
\end{figure*}

Figure \ref{fig:maxilchr} shows the
background-subtracted MAXI/GSC light curves 
in the 2--10 keV and 10--20 keV bands
from 2022 May 21 (MJD 59720) to August 29 (MJD 59820).
Each data point represents the photon flux 
in units of photons cm$^{-2}$\,s$^{-1}$ 
for each scan transit of $\sim 40$\,seconds.
They have been corrected for the effective area dependent 
on the source incident angle and the photon energy
assuming that the source has a power-law spectrum with 
a photon index $\Gamma = 2$.
The outburst continued for $\sim50$\,days.
The observed flux shows the large amplitude variation with an apparent
periodicity of about a day, as reported in \citet{Serino2022}. As we will demonstrate in Section \ref{sec:pulseAverageLC}, combined analysis of the NuSTAR and Fermi/GBM
data shows that the apparent periodicity is an alias due to the uneven
fractional sampling of the 1130\,s coherent pulsation in the MAXI/GSC
observations.

In Figure \ref{fig:maxilchr}, the hardness ratio of the 10--20 keV to
the 2--10 keV flux in 1-day time bins is plotted in the bottom panel. The
hardness peaked earlier than the flux and then gradually declined
during the outburst.

\section{Short term variability} \label{sec:timing}

\begin{figure*}
\begin{center}
\includegraphics[width=0.95\textwidth]{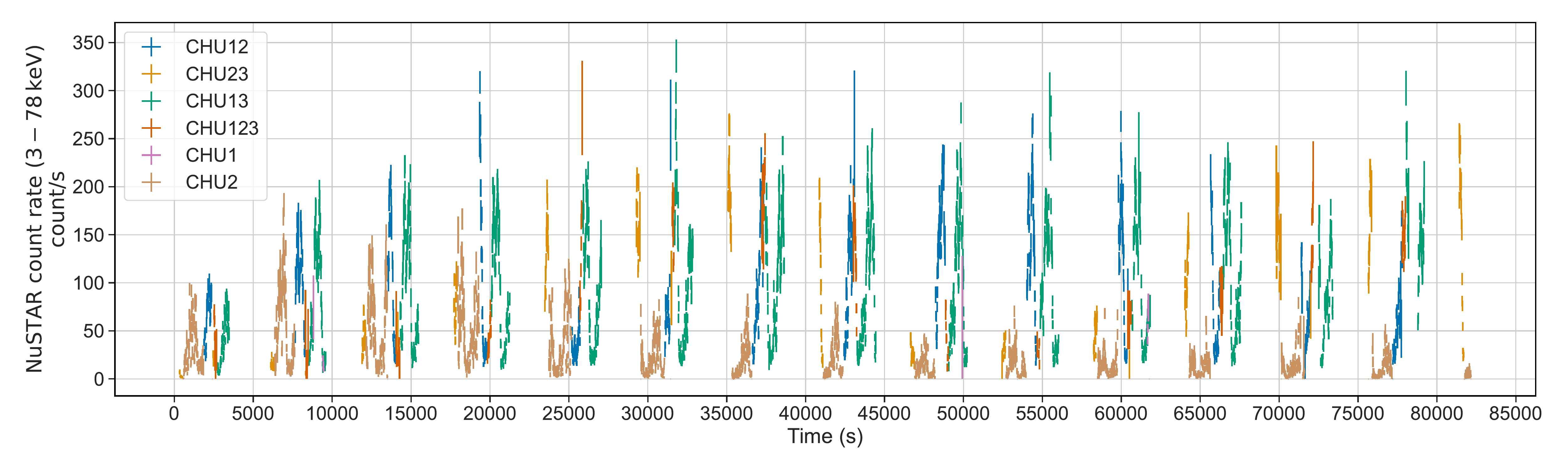}
\vfill
\includegraphics[width=0.55\textwidth]{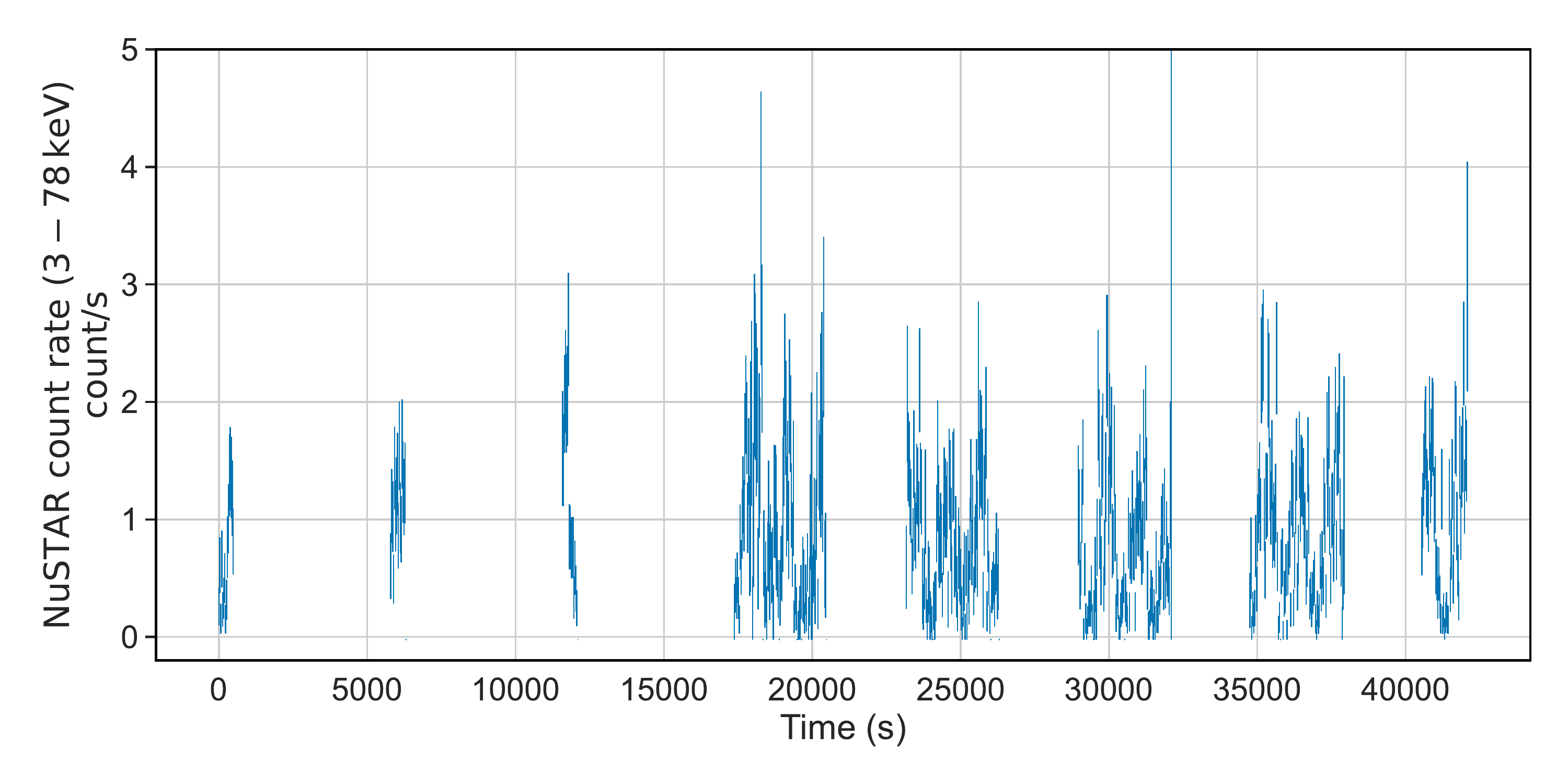}
\caption{Top: Background subtracted NuSTAR light curve (FPMA + FPMB) for observation 80801347002 (OBS1) with bin size 10\,s. The light-curve has been split into CHU combinations. We find that when only CHU2 was in use, the observed count rate dropped drastically. By inspecting of the count maps, we found that the source centroid landed less than $30^{\prime\prime}$ from the edge of the detector for CHU2, whereas for all other CHU combinations the centroid was displaced from the detector edge by about $70^{\prime\prime}$. The half-power diameter for the NuSTAR FPMs is $60^{\prime\prime}$--$70^{\prime\prime}$, meaning that the drop in observed count rate associated with CHU2 is due to proximity to the detector edge rather than intrinsic variability. Bottom: Background subtracted NuSTAR light curve (FPMA + FPMB) for observation 90801321001 (OBS2) with bin size 32\,s. Pulsations with a period of about 1100\,s can be seen in both observations.}
\label{fig:nu_lcs}
\end{center}
\end{figure*}

For both NuSTAR observations and all CHU combinations, we produced background subtracted light-curves using circular source regions with radii $45^{\prime\prime}$ and background regions with radii $90^{\prime\prime}$. The resulting light-curves, which represent the sum of FPMA and FPMB curves, are shown in Figure \ref{fig:nu_lcs}. For OBS1, we have color-coded the light-curve according to operational CHU combination. It can be seen from this figure that some CHU combinations have longer exposures than others, and each samples a different part of the NuSTAR orbit. Strong variability is clearly visible during both observations, which prompted us to perform a detailed timing analysis, including a search for and characterization of coherent pulsations.

For both observations, we extracted two lists --- one for FPMA and one for FPMB --- of cleaned and filtered source events. In the case of OBS1, rather than creating event lists for each CHU combination, we simply extracted all Mode 6 data (including CHU2) within a source region of radius $60^{\prime\prime}$ to account for the shifting centroid as well as the elevated source count rate. We note that Mode 6 does not affect clock uncertainties. For OBS2, we again used a source region of radius $45^{\prime\prime}$. We further filtered events such that those with energy outside the $3-78$\,keV energy range were excluded. We then corrected the photon arrival times for the motion of NuSTAR by shifting the times into the reference frame of the Solar System barycenter using the \code{barycorr} tool. We specified the source position in the ICRS frame using the coordinates reported by \citet{Kennea2022a}. We used the ephemeris JPLEPH.430 and the NuSTAR clockfile nuCclock20100101v147 in order to account for NuSTAR's onboard clock variations.

\begin{figure*}
\begin{center}
\includegraphics[width=0.9\textwidth]{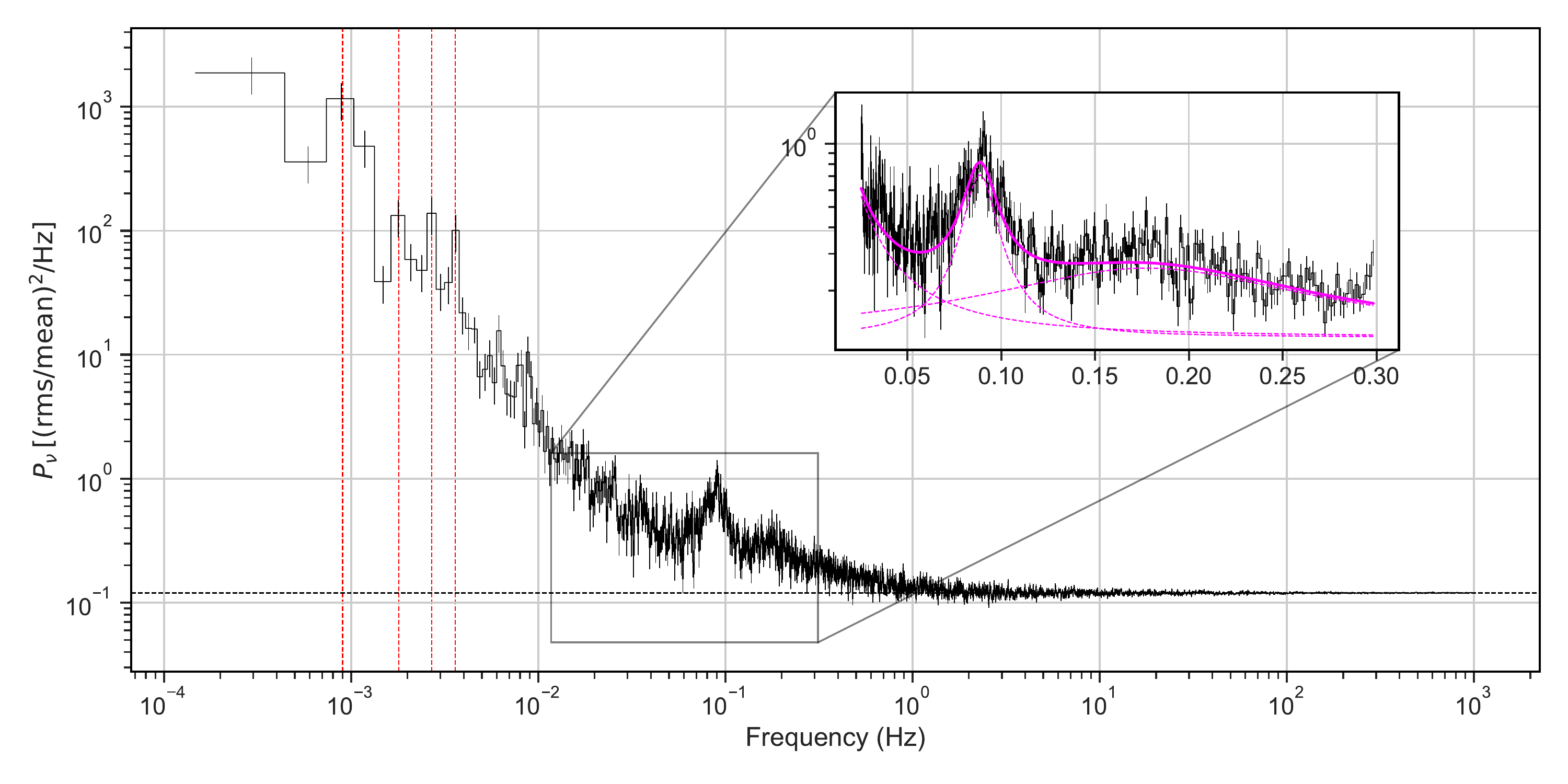}
\caption{Source normalized, averaged, FAD-corrected power density spectrum for OBS1. The power spectrum has been logarithmically rebinned for clarity. Red dashed lines are overlaid at $\nu=0.9\,\mathrm{mHz}$ and at 2, 3, and 4 times this frequency to demonstrate the robust detection of coherent pulsations in addition to harmonics. We also find evidence for quasi-periodic oscillations at a central frequency of $\nu_{0}=89\pm1\,\mathrm{mHz}$ with a harmonic at $\nu_{1}=180\pm10\,\mathrm{mHz}$, shown in the inset. The dashed magenta lines show the model components (a power law and two Lorentzians) which were fit to the data shown in the inset, and the sum of components is shown by the solid magenta line. The dashed black line at $P_{\nu}=0.12$ indicates the Poisson noise level determined by averaging all powers above $\nu=10\,\mathrm{Hz}$.}
\label{fig:FAD}
\end{center}
\end{figure*}

With energy- and region-filtered, barycenter-corrected event lists in hand, we proceeded to produce power spectra. Due to the low count rate and short exposure time of OBS2, we only present power spectra for OBS1. We split the observation into 9 segments of length 3400\,s each, avoiding orbital gaps. For each of these segments we binned the events into light-curves with bin size $2048^{-1}$\,s. For each light-curve, we calculated the power density spectrum (PDS). Because the deadtime after each NuSTAR event can imprint significant timing artifacts, even for moderate count rates \citep{Bachetti2015}, we specifically produced the Fourier Amplitude Difference (FAD) corrected PDS \citep{Bachetti2018} for each light-curve using Stingray. This method takes advantage of the simultaneous observation of the source with two independent detectors, as in the case of the two FPMs on NuSTAR, in order to filter out the effects of deadtime. We then averaged the 9 FAD-corrected power spectra and source-normalized the resulting PDS according to the ratio of source to background count rates. The final, rms-normalized PDS is shown in Figure \ref{fig:FAD}, which has been rebinned logarithmically for legibility. Four narrow peaks are visible at multiples of $\nu=0.9$\,mHz, demonstrating clear evidence for coherent pulsations with harmonics. A wider feature resembling a signature of quasi-periodic oscillations (QPO) can be seen at higher frequencies, around $\nu=0.09$\,Hz, along with a smaller bump around twice this frequency.

\subsection{Quasi-periodic oscillations at 89 mHz}\label{sec:qpo}

\begin{deluxetable}{cccc}
\tablenum{2}
\tablecaption{QPO parameters for NuSTAR observation 80801347002 with 90\% confidence intervals. \label{tab:qpo_params}}
\tablewidth{0pt}
\tablehead{\colhead{Component} & \colhead{$\nu_0$ (mHz)} & \colhead{$Q$} & rms (\%)}
\startdata
\noalign{\smallskip}
Fundamental   &   $89 \pm 1$        &   $4.6 \pm 1.0$   & $13.4 \pm 1.1$ \\
2nd Harmonic      &   $180 \pm 10$      &   $1.2 \pm 0.3$   & $18.5 \pm 2.2$ \\
\noalign{\smallskip}
\hline
\enddata
\end{deluxetable}

In order to characterize the quasi-periodic variability, we fit the PDS bins between 0.03\,Hz and 0.3\,Hz to a model consisting of a power law continuum plus two Lorentzian features, which are frequently used to model QPOs \citep{Belloni2002}. The power, $P(\nu)$, of the Lorentzian components can each be written

\begin{equation}
    P(\nu) = \frac{r^2\Delta}{\pi}\frac{1}{\Delta^2 + \left( \nu - \nu_0 \right)^2}
\end{equation}

\noindent where $r$ is the integrated fractional rms under the Lorentzian, $\Delta$ is the half-width at half-maximum (HWHM), and $\nu_0$ is the centroid frequency of the component. We may further define the ``coherence" of the feature, $Q \equiv \nu_0/2\Delta$, which represents a measure of the relative width of the Lorentzian. The results of this model fitting are shown in Table \ref{tab:qpo_params}. We found that the model describes the features well, yielding centroid QPO frequencies of $89\pm1$\,mHz and $180\pm10$\,mHz and a broad band noise component with a power law index of $-2.0\pm0.4$ (consistent with the tail of a low-frequency Lorentzian component). We determined the errors on these values using the covariance matrix returned by the SciPy \citep{Scipy} least squares fitting function \code{curve\_fit}. Because the higher-frequency component is consistent with twice the centroid frequency of the lower-frequency component, we conclude that they represent a fundamental and second harmonic of the same process. We discuss the physical nature of this timing feature in Section \ref{sec:qpobfield}.

In order to determine whether or not the QPO may have been present during OBS2, we similarly produced a FAD-corrected PDS for this observation, this time using segments of length $256$\,s and a light-curve binning resolution of $128^{-1}$\,s. Visual inspection indicated that the PDS in the resulting frequency range is consistent with Poisson noise. We attempted to place an upper limit on the strength of the QPO by least-squares fitting a Lorentzian component to the PDS in the frequency range $0.01$\,Hz--$1$\,Hz. After determining the best-fit parameters, we increased the amplitude of the Lorentzian in steps of 0.01\% of the best-fit value until the resulting $\chi^2$ statistic had increased by $6.635$. We thereby placed a 99\% confidence upper limit on the QPO rms of 27\%. This upper limit, which is significantly higher than the rms observed during OBS1, demonstrates that the data are simply insufficient to detect the QPO during OBS2.

\subsection{Pulsations at 1100 s} \label{sec:pulsations}

\begin{figure*}
\begin{center}
\includegraphics[width=0.49\textwidth]{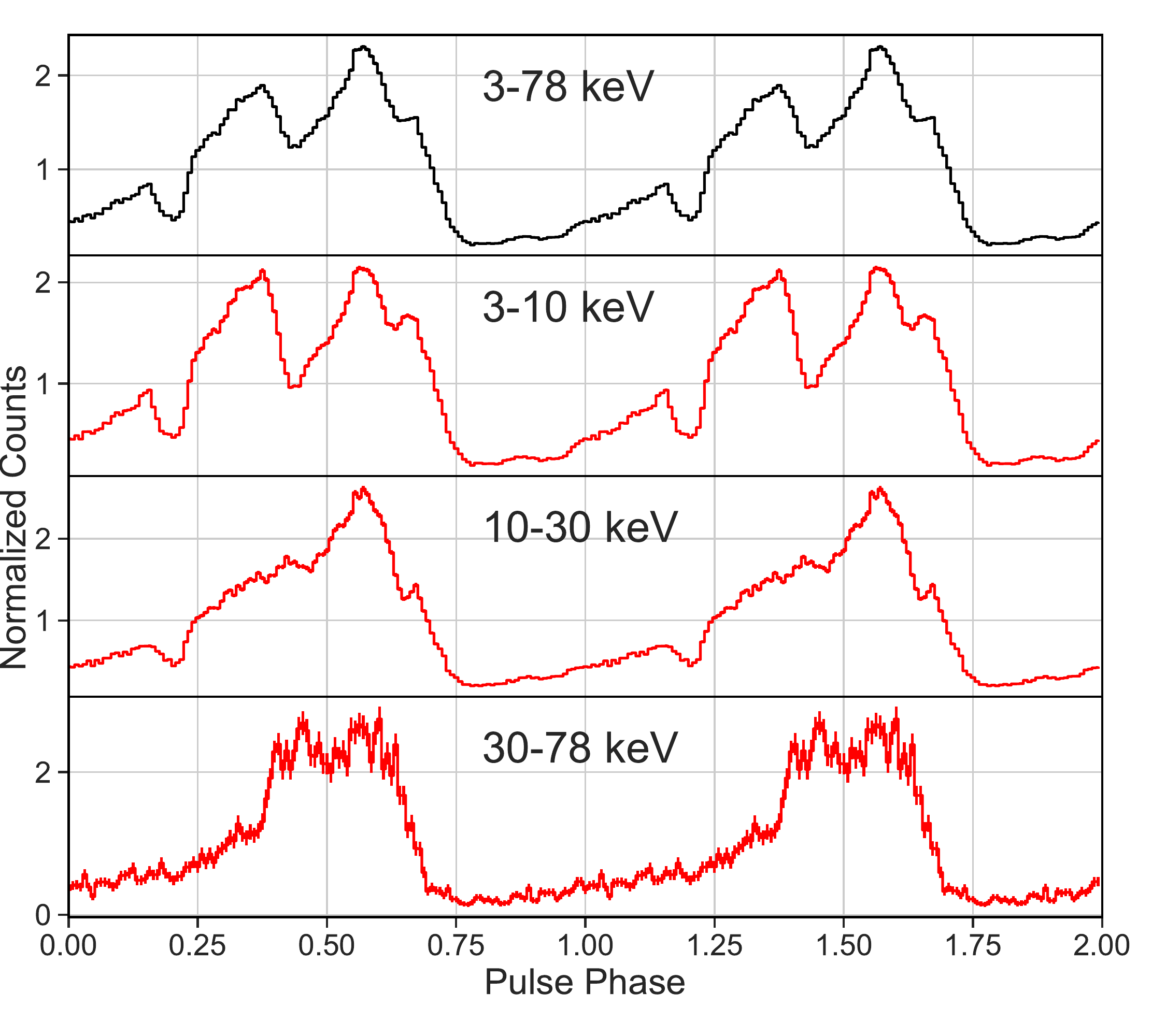}
\hfill
\includegraphics[width=0.49\textwidth]{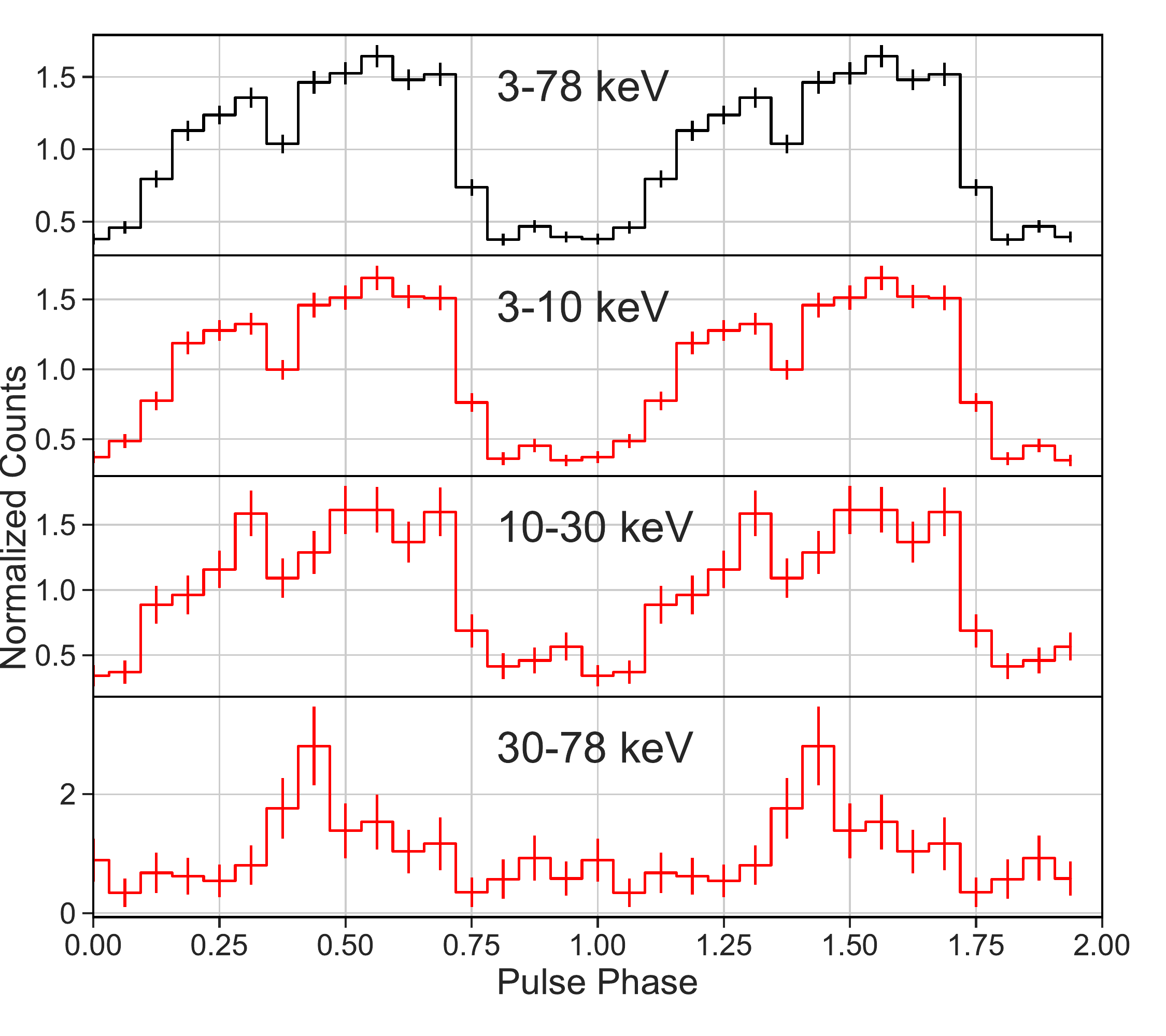}
\caption{Pulse profiles constructed for NuSTAR observations 80801347002 (left) and 90801321001 (right). The pulse profile for the full NuSTAR energy band is shown in black in the top panels, and the pulse profiles for narrower energy bins are shown in red in the lower three panels of each figure. While the strength of pulsations does not vary much for different energy bins, there is clear energy-dependence in the shape of the profiles. During both observations, the pulse profile has multiple peaks at lower energies but at higher energies the shape of the profile changes dramatically.}
\label{fig:pulse_profiles}
\end{center}
\end{figure*}

In order to further constrain the pulse period during OBS1, we calculated the $Z^2_n$ statistic \citep{Buccheri1983} with 4 harmonics ($n=4$) for 1000 linearly spaced spin frequencies between $735\,\mathrm{\mu Hz}$ and $1029\,\mathrm{\mu Hz}$. Using SciPy, we produced a cubic spline interpolation of the $Z^2_4$ distribution in order to determine the pulse frequency corresponding to the maximum statistic and the respective error region. Because the $Z^2_n$ statistic follows the same probably density function as $\chi^2$ with $2n$ degrees of freedom, we defined the $90\%$ confidence regions according to the change in pulse frequency which corresponds to a change in the statistic of $\Delta Z^2_4 = 13.36$. Using this method, we found a pulse period of $P=1129.09 \pm 0.04$\,s. We performed the same $Z^2_4$ search for OBS2, searching the same frequency range and found that the spin period had decreased significantly to a value of $P=1085\pm1$\,s. We confirmed that these measurements indeed represent the fundamental spin frequency by performing a similar search using the OBS1 events, this time calculating the statistic for one harmonic ($Z^2_1$) over all frequencies less than $1900\,\mathrm{\mu Hz}$. We found that the signal at $1130$\,s was still clearly visible, and we found no peaks of comparable magnitude in this range other than those corresponding to integer multiples of the orbital period of NuSTAR. Furthermore, we found that when performing this same search with more harmonics, up to $n=4$, the peak at 1130\,s remained the strongest aside from those corresponding to the NuSTAR orbit. We therefore confirm that we have measured the fundamental spin frequency rather than a harmonic.

The neutron star experienced a secular spin-up of around $\dot{P}_\mathrm{sec}\approx-10^{-5}\,\mathrm{s\,s^{-1}}$ between NuSTAR observations. As we will show in Section \ref{sec:pulseAverageLC}, Fermi GBM measurements of the pulse period throughout the outburst indicate that the long-term spin-up rate during the 6-day interval including OBS1 was $\dot{P}\approx-7\times10^{-6}\,\mathrm{s\,s^{-1}}$. In order to estimate the spin-up rate throughout OBS1, we split the observation into 2 intervals of approximately equal length and measured the spin period for each interval via the same method we used to measure the spin period for the full observations. We found that over the course of 38\,ks, the spin period decreased from $1129.1\pm0.1$\,s to $1128.4\pm0.1$\,s, resulting in a spin-up rate of $\dot{P}=(-1.6\pm0.4)\times10^{-5}\,\mathrm{s\,s^{-1}}$. This value is twice the inferred secular rate, indicating short-term variability in the rate of spin-up, which may be stochastic in nature or may be attributed in part to orbital motion of the pulsating neutron star. Using the same method, we were unable to constrain the spin-up rate experienced by the neutron star throughout OBS2, instead placing a $99\%$ upper limit of $|\dot{P}|<6\times10^{-4}\,\mathrm{s\,s^{-1}}$. The results of our pulsation analysis are tabulated in Table \ref{tab:pulse_params}. We further discuss the evolution of the spin period and its implications for the magnetic field of the neutron star in Section \ref{sec:ghoshandlamb}.

\begin{deluxetable*}{ccccc}[tbh]
\tablenum{3}
\tablecaption{Pulsation parameters for both NuSTAR observations with 90\% confidence intervals (upper limit given at 99\% confidence). The reference time, $T_{\mathrm{ref}}$, is given in the barycentric reference frame of the Solar system. Pulse fractions (PF) are given for the full 3-78\,keV energy range. \label{tab:pulse_params}}
\tablewidth{0pt}
\tablehead{\colhead{OBSID} & \colhead{$T_{\mathrm{ref}}$ (MJD)}     & \colhead{$P$ (s)} & \colhead{$-\dot{P}$ (s/s)} & \colhead{PF (\%)}}
\startdata
\noalign{\smallskip}
80801347002 & $59751.93620360$   &  $1129.09\pm0.04$   &  $(1.6\pm0.4)\times10^{-5}$ & $84.4 \pm 0.7$  \\
90801321001 & $59805.84542452$   &  $1085\pm1$  &  $<6\times10^{-4}$ & $63 \pm 6$  \\
\noalign{\smallskip}
\hline
\enddata
\end{deluxetable*}

Having determined the spin period of the source during both NuSTAR observations, we used \code{nuproducts} to produce livetime-corrected light-curves with 1-second binning for the full NuSTAR bandpass as well as for three smaller energy bins: 3--10\,keV, 10--30\,keV, and 30--78\,keV. We folded each of these light-curves into pulse profiles, shown in Figure \ref{fig:pulse_profiles}, according to the pulse period we extracted for each observation. From these profiles we calculated the pulse fraction, defined as $PF \equiv (I_\mathrm{max}-I_\mathrm{min})/(I_\mathrm{max}+I_\mathrm{min})$, where $I_\mathrm{max}$ and $I_\mathrm{min}$ are the maximum and minimum values of the pulse profiles, respectively. The emission is highly pulsed, with a pulse fraction of $PF=84.4\pm0.7\%$ during OBS1 and $PF=63\pm6\%$ during OBS2. During both observations, we find a small increase in the pulse fraction with photon energy. We list the pulse periods and pulse fractions for both NuSTAR observations in Table \ref{tab:pulse_params}.

The pulse profiles show significant deviations from a simple sinusoid (this is particularly clear for OBS1 which has significantly more counts) and the structure of the pulsations differs drastically with photon energy. At low energies (3--10\,keV) we see 3 to 4 separate peaks in the profile, while for photon energies above 10\,keV, the number of distinct peaks decreases. For the hardest X-rays available to NuSTAR, a single spiky plateau is observed. The combined effect is a soft peak in the phase interval $0.2<\phi<0.4$ and a hard peak for $0.4<\phi<0.7$. This rich behavior is likely due to the changing viewing angle of the complex magnetic field as the neutron star rotates, resulting in an energy-dependent beam structure that may have both a fan-like and pencil-like component \citep{Iwakiri2019}. Future work may include physical modeling of the energy-dependent pulse profile, combined with pulse-phase spectroscopy, which will help illuminate the structure of the neutron star magnetic field as well as the interaction of the field with accreting material.

\section{Spectral modeling and variability} \label{sec:spectra}

\begin{figure*}
\begin{center}
\includegraphics[width=0.49\textwidth]{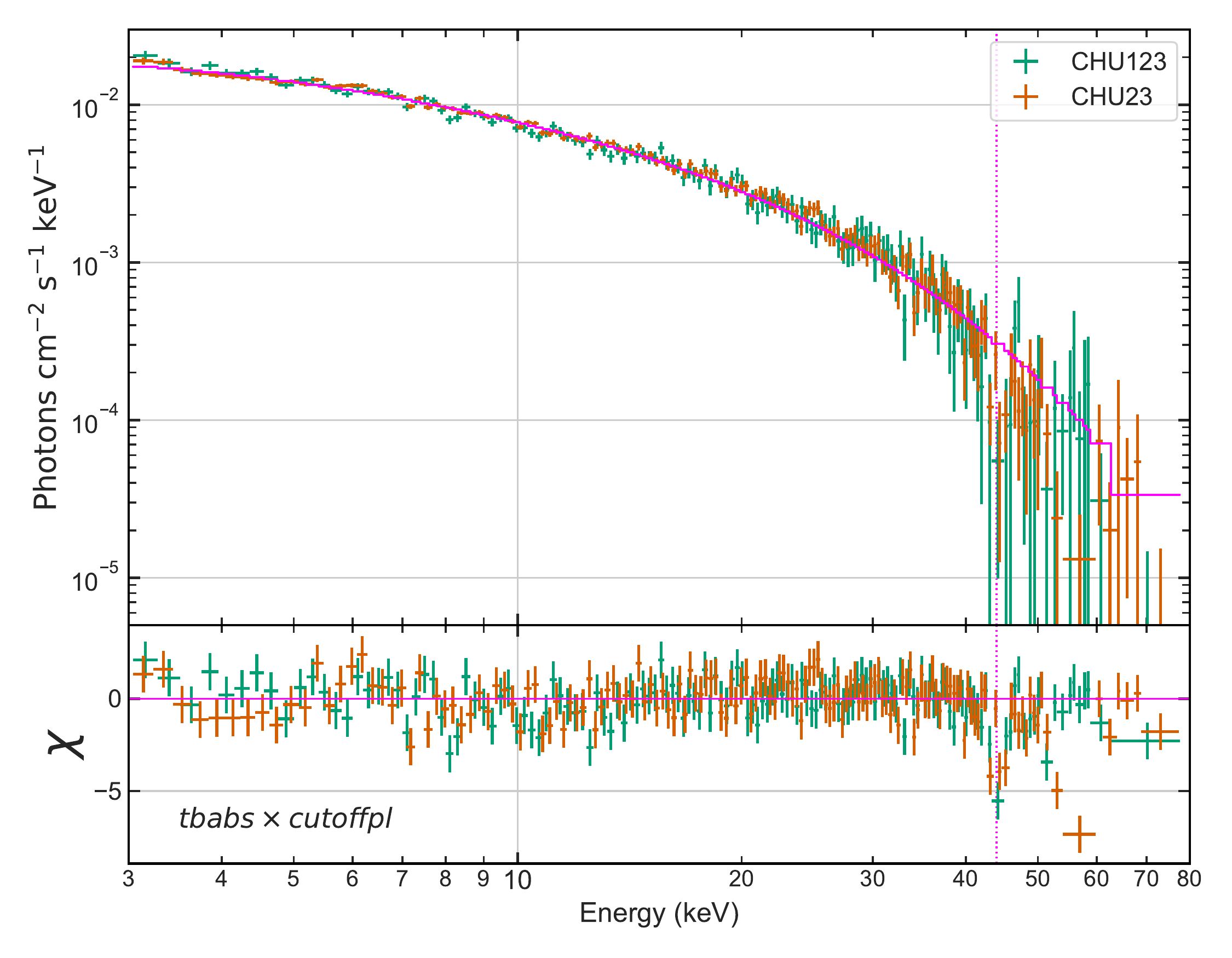}
\hfill
\includegraphics[width=0.49\textwidth]{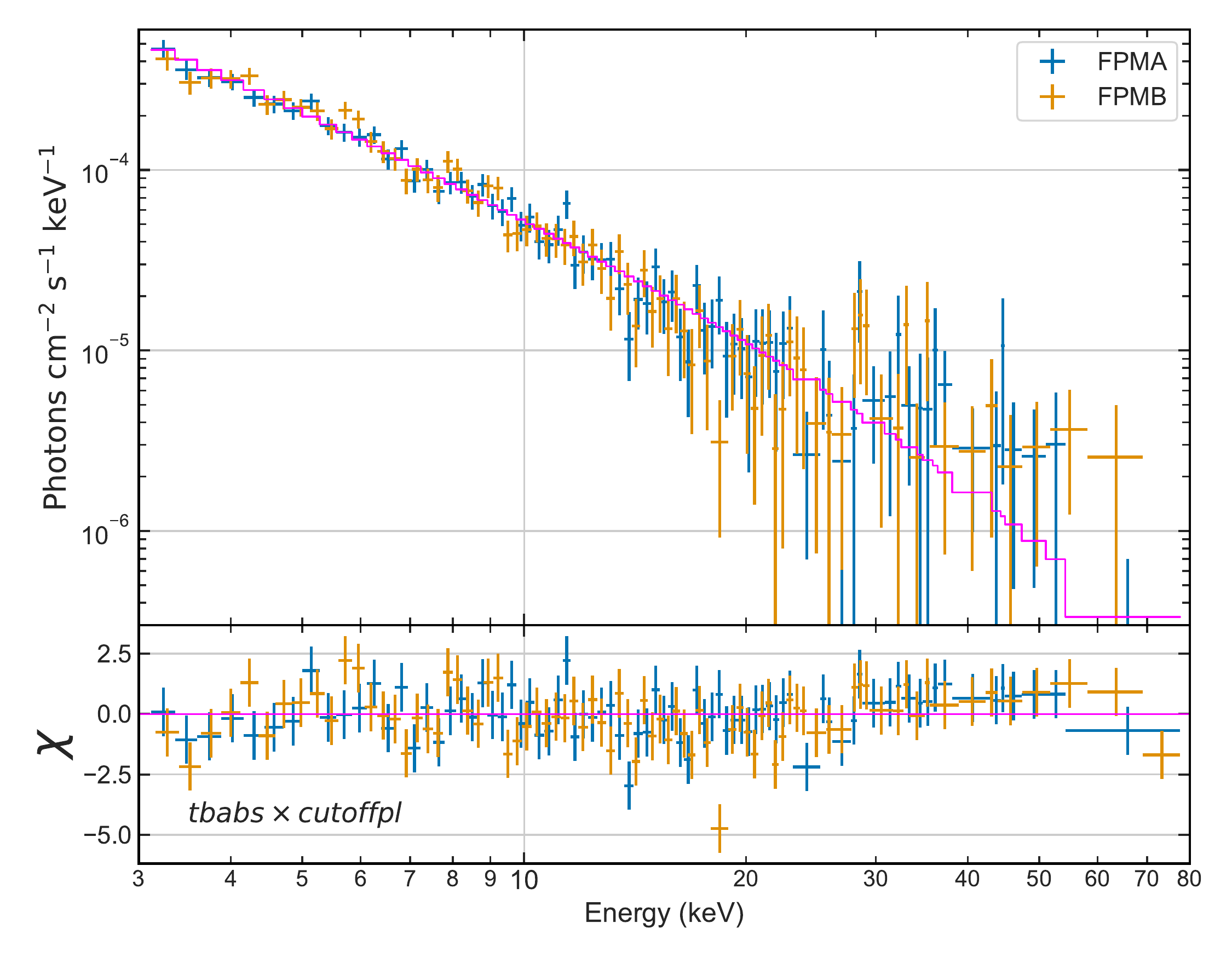}
\caption{NuSTAR spectra, models, and residuals for observations 80801347002 (left) and 90801321001 (right). For observation 80801347002 we simultaneously fit the spectra measured by FPMB in Mode 6 for  CHU combinations CHU123 and CHU23. Because the source fell on the gap between detectors on FPMA, we excluded the spectra measured by FPMA. For observation 90801321001, we show the spectra as measured by both focal plane modules in the usual Mode 1 (Science Mode). For both observations, the spectrum is described well by an absorbed power law with a high energy cutoff. We show the best-fit models in magenta. The vertical dashed line at 44\,keV on the left figure indicates the tentative absorption feature.}
\label{fig:nu_spectra}
\end{center}
\end{figure*}

In order to characterize the source spectrum during OBS1, we chose to model only data from FPMB using CHU combinations CHU23 and CHU123. We chose these data due to their clear, single-peaked source centroid visible in DS9, and because the source centroid was far enough from the chip gap that it could be avoided during source extraction. We produced spectra from each of the two event lists, and we performed a joint fit, tying all model parameters together aside from a multiplicative factor which was set to unity for CHU123 and allowed to vary for CHU23. We extracted the  spectra using circular source and background regions with radii $45^{\prime\prime}$ and $90^{\prime\prime}$, respectively. For OBS2, we followed the same procedure, but rather than using only FPMB, we extracted spectra for both FPMA and FPMB, and there was no need to split the events according to CHU combination. The source spectra as well as the residuals resulting from spectral modeling are shown in Figure \ref{fig:nu_spectra}. For both obervations, the spectra were binned using the optimal binning procedure described by \cite{Kaastra2016}. For the sake of visual clarity, we further binned spectra such that each bin shown in Figure \ref{fig:nu_spectra} has a significance of 1-$\sigma$. During spectral fitting, we considered data across the entire NuSTAR bandwidth of 3--78\,keV, but we note that for OBS2 the background contributes at about the same level as the source above about 50\,keV. For the purpose of model fitting, we used the W-statistic, which is a modified version of the Cash statistic \citep{Cash1979} that takes into account the contribution of background counts.

We found that the spectrum during both observations could be described relatively well using a simple phenomenological model consisting of an absorbed power law with a high-energy cutoff. We used the Xspec model \code{cutoffpl} to model the power component. This model can be written

\begin{equation}
    F(E) = KE^{-\alpha}\exp{(-E/\beta)}
\end{equation}

\noindent where $F(E)$ is the photon flux per keV at energy $E$, $\alpha$ is the power law photon index, $\beta$ is the e-folding energy of the high-energy cutoff, and $K$ is a normalization factor. For both spectra, we modeled absorption using the Xspec model \code{tbabs}, with abundances provided by \citet{Wilms2000}. Using the FTOOL command \code{nh}, we derived a Galactic hydrogen column density in the direction of \protag\ of $5.5\times10^{21}\,\mathrm{cm^{-2}}$ from the HI4PI full-sky HI survey \citep{HI4PI}. Our fits were not sensitive to the column density, and we therefore froze the parameter at $n_\mathrm{H}=5.5\times10^{21}\,\mathrm{cm^{-2}}$ for both observations.

In order to estimate the confidence regions for each parameter, we used the built-in Xspec command, \code{error} to vary each parameter in the positive and negative directions until the fit statistic increased by 2.706, corresponding to the 90\% confidence region. In addition to the spectral parameters, we calculated the unabsorbed, bolometric flux using the Xspec model \code{cflux}. We set the energy range of the model to $0.1-100$\,keV, and we similarly set the energy bin array for flux integration to the same range with 1000 logarithmically spaced bins using the Xspec command \code{energies}. The latter is a necessary step to accurately integrate flux outside of the NuSTAR energy range.

\begin{deluxetable*}{cccc}
\tablenum{4}
\tablecaption{Continuum spectral parameters determined using Xspec. Parameters marked with a dagger ($\dagger$) were frozen during parameter estimation. \label{tab:spectral_params}}
\tablewidth{0pt}
\tablehead{
\colhead{Model component}                       &   \colhead{Parameter}     & \colhead{Obs1} & \colhead{Obs2}}
\startdata
\noalign{\smallskip}
\code{tbabs}                                    & $N_{\rm H}$ ($10^{22}\,\mathrm{cm}^{-2}$)         & $0.55^\dagger$       & $0.55^\dagger$             \\
\noalign{\smallskip}
\hline
\multirow{3}{*}{\code{cutoffpl}}    & $\alpha$ 			    & $0.27 \pm 0.04$           & $1.78\pm 0.2$    \\
                                    & $\beta$ (keV)         &  $12.1^{+0.5}_{-0.4}$     & $31^{+71}_{-14}$            \\
                                    & Norm ($10^{-3}\mathrm{\frac{photon}{keV\,cm^2\,s}}$)    & $33 \pm 2$                & $4.4^{+0.2}_{-0.1}$      \\
\noalign{\smallskip}
\hline
\hline
\noalign{\smallskip}
\multicolumn{2}{c}{$F_\mathrm{bol}(10^{-11}\,\mathrm{erg\,cm^{-2}\,s^{-1}})$\tablenotemark{\small a}}   & $361\pm7$           & $4.2^{+1.2}_{-0.8}$             \\
\multicolumn{2}{c}{$L_\mathrm{bol}(10^{34}\,\mathrm{erg\,s^{-1}})$\tablenotemark{\small b}}             & $560^{+150}_{-100}$    & $6.5^{+2.6}_{-2.2}$             \\
\multicolumn{2}{c}{$W/\mathrm{d.o.f.}$}                                                                 & 399/357 & 304/290                         \\
\noalign{\smallskip}
\enddata

\tablenotetext{\small a}{\footnotesize Unabsorbed bolometric ($0.1-100\,\mathrm{keV}$) flux.}
\tablenotetext{\small b}{\footnotesize Unabsorbed bolometric ($0.1-100\,\mathrm{keV}$) luminosity assuming isotropic emission at a distance of $3.6^{+0.3}_{-0.2}\,\mathrm{kpc}$. Uncertainties include the contribution due to uncertainty in the distance measurement.}
\end{deluxetable*}

The resultant spectral parameters, bolometric flux, and luminosity for each observation are listed in Table \ref{tab:spectral_params}. In addition to dropping in flux by nearly two orders of magnitude, the source underwent a clear change in spectral state between the two NuSTAR observations. While the power law steepens drastically and the cutoff energy increases, both spectra yield spectral parameters which are typical of accreting HMXBs \citep{Coburn2002}.

Significant residuals can still be seen in the spectrum of OBS1 for photon energies above 40\,keV (see Figure \ref{fig:nu_spectra}, left panel). In particular, we noted a feature resembling an absorption line around 44\,keV (marked by the vertical dashed magenta line in Figure \ref{fig:nu_spectra}) which appeared in both CHU combinations which we investigated, unlike the residuals seen only in CHU23 above 50\,keV. Given that cyclotron resonance scattering features (CRSF) are often detected in the spectra of BeXRBs at energies of tens of keV \citep{Staubert2019}, we attempted to model this absorption feature using the multiplicative cyclotron scattering component \code{cyclabs} in Xspec. We set the depth of the second harmonic to 0 since a feature at $\sim90$\,keV would not be visible to NuSTAR. We found that when we left the centroid, the width, and the depth of the first harmonic free, the data preferred an absorption feature centered at 53\,keV with width 13.6\,keV, while the narrower feature at 44\,keV remained visible in the residuals. This indicated to us that attempts to characterize the absorption feature are sensitive to the continuum model at the high end of the NuSTAR bandpass. A wide absorption feature like this may simply indicate that the single cutoff power law model does not perfectly describe the hard X-ray continuum. To try to better understand the narrow component, we therefore fit the model while freezing the centroid energy at a value of 44\,keV. This resulted in an improvement to the fit statistic of $\Delta W=-34$ and a decrease of 2 in the degrees of freedom. The resulting best-fit width of the feature was around 2\,keV, and the optical depth was around unity. We roughly estimated the significance of the feature by calculating the ratio of the best-fit value of the optical depth to the error of that parameter. This method yields a significance of $2.7\sigma$. As a check, we also produced spectra using background regions on a different detector, but the residual features were not significantly affected. We discuss the implications of this spectral feature and several important caveats in Section \ref{sec:crsf}.


\section{Reconstructing the pulse-averaged outburst} \label{sec:pulseAverageLC}

\begin{figure}
\begin{center}
\includegraphics[width=0.5\textwidth]{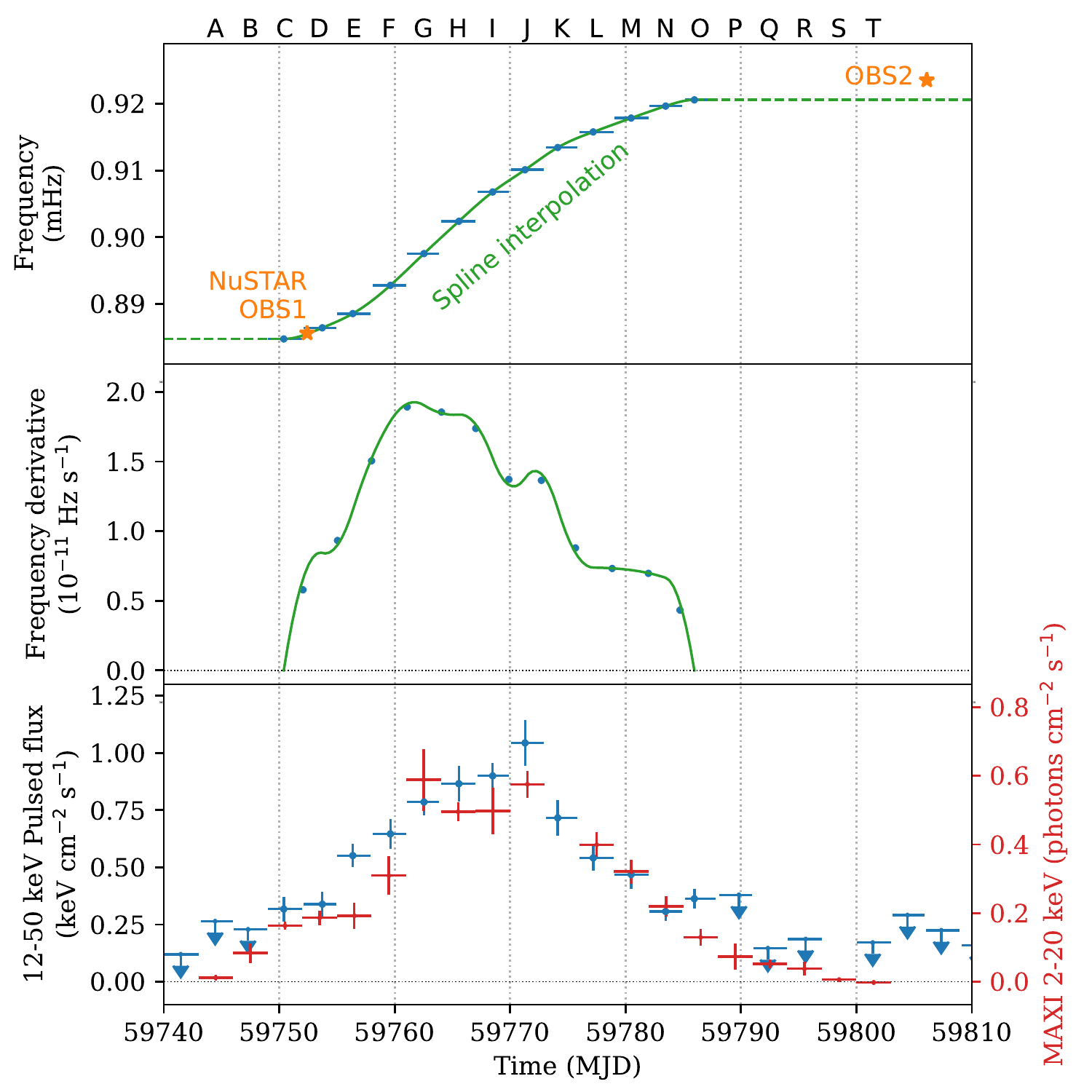}
\caption{
  \protag\  pulsed emission history during the present
  outburst obtained from the Fermi/GBM pulsar data. 
  {\bf Top:} Barycentric pulse frequency measured during each 3 day
  interval, labeled A, B, ..., T as shown above the top axis. Data obtained
  from the two NuSTAR observations (in Table \ref{tab:pulse_params})
  are also plotted as orange stars. The model function
  approximating the continuous change with the spline interpolation is plotted as a solid green.
  {\bf Middle:} Pulse-frequency derivative calculated from adjacent pulse period
  measurements (blue) and that of the spline-interpolation model (green) {\bf
    Bottom:} 12--50\,keV pulsed flux reported by Fermi/GBM (blue). The pulse-phase-average 2--20\,keV
  total flux obtained from the MAXI/GSC data (also shown in Figure
  \ref{fig:maxilcave}) is shown in red.
}
\label{fig:fgbmhist}
\end{center}
\end{figure}

\begin{figure}[tbh]
\includegraphics[width=0.49\textwidth]{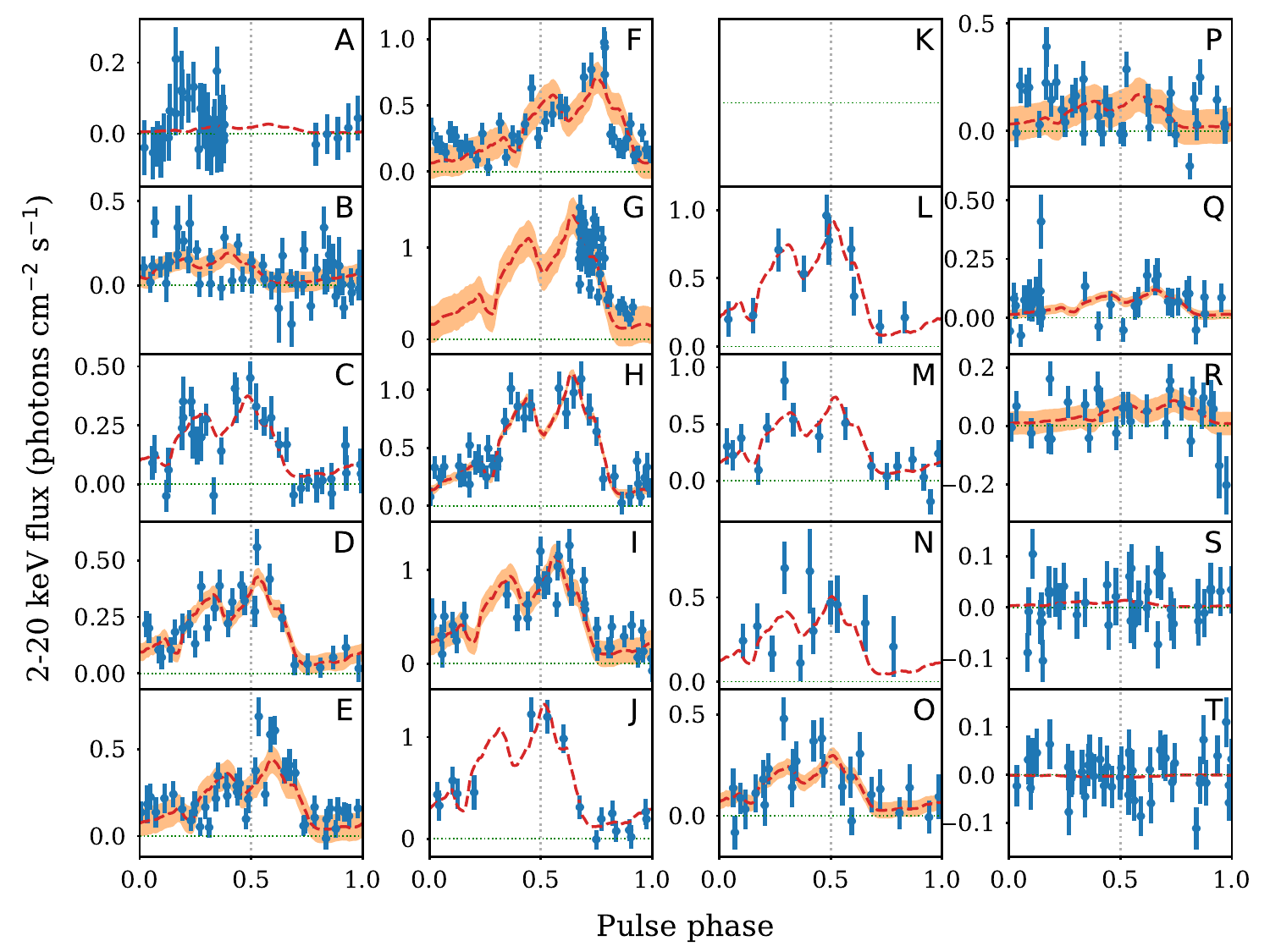}
\caption{Light curve of the MAXI/GSC scan data (blue points) folded at the pulse
  period obtained from the Fermi/GBM pulsar data and
  the best-fit pulse-profile model with a template extracted from the
  NuSTAR data (red dashed line) for each 3 day interval, A, B, ...,
  T (see Figure
  \ref{fig:fgbmhist}). The source was not observed by MAXI/GSC during interval K. Orange-shaded regions represent the systematic
  error added in quadrature to the statistical error during model fitting. }
\label{fig:maxilcfit}
\vspace{5mm}
\includegraphics[width=0.49\textwidth]{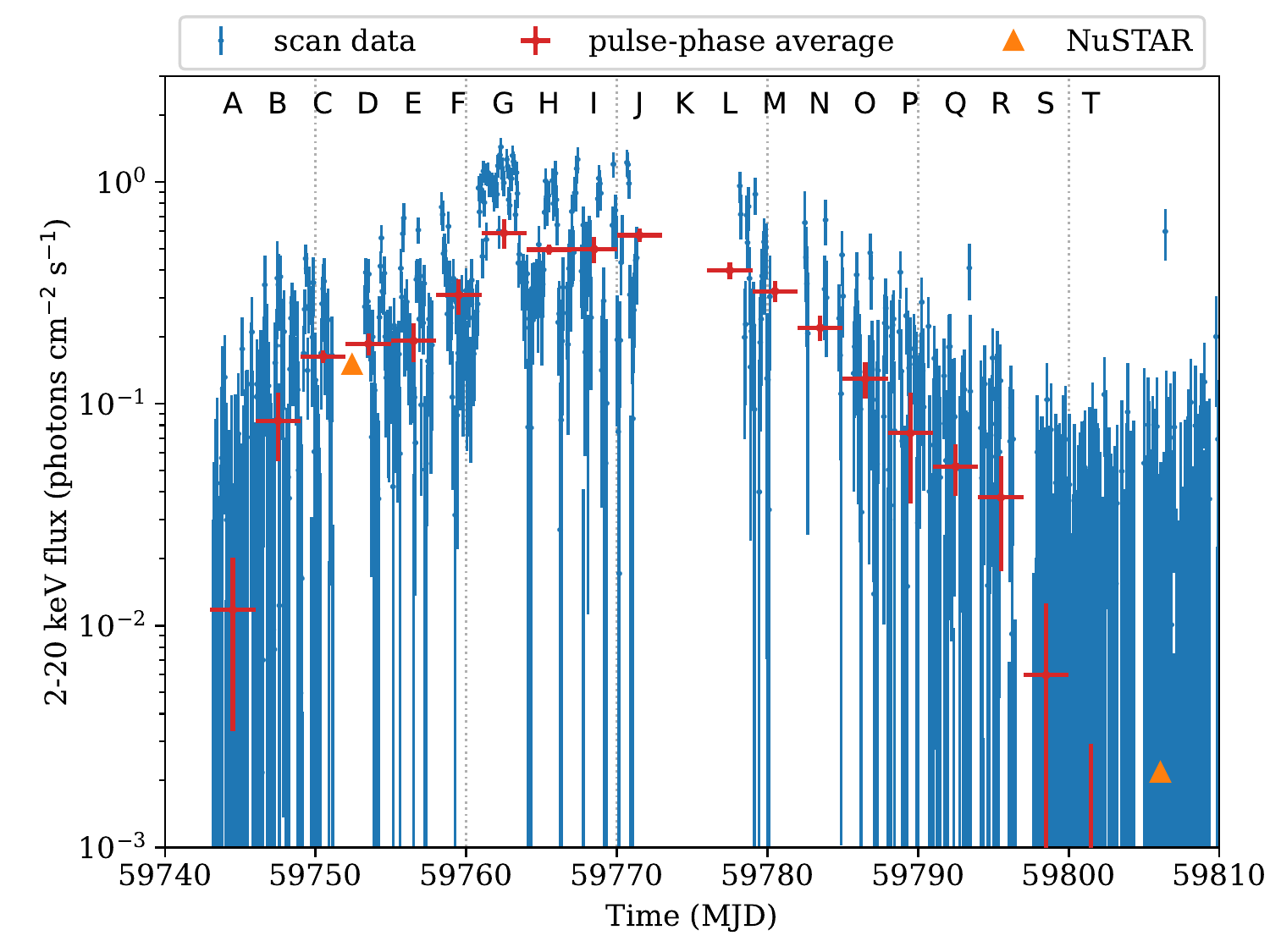}    
\caption{ MAXI/GSC 2--20\,keV light curves extracted from the raw scan data (blue)
  and from the pulse-phase-average for each 3 day interval calculated
  using the best-fit pulse-profile models in Figure
  \ref{fig:maxilcfit} (red). 
  NuSTAR 2--20 keV fluxes estimated from the best-fit spectral models are also shown as orange triangles.}
\label{fig:maxilcave}
\end{figure}

The MAXI light-curve exhibits dramatic variability on the timescale of days. However, we have shown that the source pulses with a period of around 1100\,s --- only a factor of a few smaller than the orbital period of the ISS, which is around 90\,minutes --- with a considerable pulse fraction. Thus, it is conceivable that the variability observed in the MAXI light-curve is not due to actual days-timescale variability but rather is due to variability in the overlap of pulsations with individual MAXI exposures. Additionally, the pulse period evolution reported by Fermi GBM demonstrates a remarkably smooth spin-up throughout the outburst, at least on several-day timescales (see Figure \ref{fig:fgbmhist}, top panel). This provides further impetus to determine whether the variability in the MAXI light-curve is intrinsic or artificial.

To study the time evolution of the pulsating X-ray emission throughout the outburst, we performed pulse-phase-resolved analysis of the
MAXI/GSC data.
%
Pulsation periods of bright X-ray binary pulsars in our Galaxy
including \protag\, have been continuously monitored by the
Fermi/GBM \citep{2020ApJ...896...90M}. For this analysis we utilized the data
archived on the website of the Fermi/GBM pulsar project.\footnote{https://gammaray.nsstc.nasa.gov/gbm/science/pulsars.html}

Figure \ref{fig:fgbmhist} shows the observed time variations of
the pulse frequency, frequency derivative, and
pulsed X-ray flux measured with the Fermi/GBM 
during the present outburst.
%
When the source was bright enough, the pulse frequency was determined
every 3 days. We therefore label twenty 3-day intervals between MJD 59743 and
59793: A, B, C, …, T. These labels and the intervals which they correspond to are shown in Figure \ref{fig:fgbmhist}. Fermi/GBM recorded the pulse frequency for intervals C--O. We note that MAXI/GSC did not observe the source during interval K.
The pulse frequency increased monotonically according to the X-ray
intensity. This suggests that the pulsar spun up due to mass
accretion onto the neutron star.
%
%
For each pair of adjacent pulse frequency measurements by Fermi/GBM, we determined the intervening first frequency derivative by performing a linear fit of the two frequencies.
The obtained frequency derivative evolution, shown in the middle panel of Figure \ref{fig:fgbmhist}, appears to correlate well with the
pulsed X-ray emission, shown in the bottom panel of the same figure.

We next produced folded pulse profiles using the MAXI/GSC data.
For each GSC scan transit, a given target object on the sky is observed
for $\sim 40$\,s. Compared to the \protag\ pulse period of $\sim
1100$\,s, the scan duration is so short that it can be considered
instantaneous.
%
%
%
To calculate the coherent pulse phase at every scan transit throughout
the outburst, we constructed a model function of the continuous
pulse-frequency change $\nu(t)$ via spline interpolation of the
Fermi/GBM data. The obtained model function is plotted along with the data in
Figure \ref{fig:fgbmhist}.
Using $\nu(t)$, the sequential pulse phase $\phi(t)$ at a time $t$ is
expressed as
\begin{equation}
  \phi(t) = \int_{t_0}^{t}\nu(\tau)d\tau  
  \label{equ:pulsephase}
\end{equation}
where 
$t_0$ is
the phase-zero epoch, i.e., $\phi(t_0)=0$.
We defined the phase-zero epoch to be the same as 
that in the timing analysis of OBS1
so that obtained pulse profiles 
are expected to be consistent
with that of the NuSTAR analysis results.

Using the pulse phase $\phi(t)$ calculated from the modeled $\nu(t)$,
we extracted folded light curves of the GSC 2--20 keV data. Figure
\ref{fig:maxilcfit} shows the folded light-curves we obtained for each of the
twenty 3-day intervals which we defined.
During the bright period when the source photons were significantly
detected, i.e. the intervals C--O, flux modulation with respect to pulse phase was clearly observed.
This reveals that the large amplitude swings in the raw GSC-scan light
curves in Figure \ref{fig:maxilchr} are due to the large pulsed
fraction.
Although the phase coverage is rather sparse in several intervals, the
folded profiles in Figure \ref{fig:maxilcfit} appear to be largely
consistent with the profile we produced using NuSTAR.

We then fit all the obtained profiles with a model employing the
pulse profile of OBS1 (Figure
\ref{fig:pulse_profiles}) as the template and introducing two free
parameters, normalization factor and phase-0 offset. In Figure
\ref{fig:maxilcfit}, the best-fit pulse-profile models are overlaid on
the data.
In most of the 3-day data intervals,
the model describes the data fairly well. However, for some intervals,
the best-fit $\chi^2$ values are not within the 90\% confidence limits
of the statistical uncertainties.
This likely represents systematic errors not included in the
fitting model such as the dependence of the pulse profile shape on the
outburst phase and energy band as well as errors on the pulse-phase
model $\phi(t)$.
We thus repeated the model fit, gradually adding a systematic error on the
model (added in quadrature with the statistical error on the data) until the fit became acceptable within the 90\%
confidence level.
Using this method we found that the required systematic errors were at most 10\%,
which is smaller than the statistical error for each scan.
In Figure \ref{fig:maxilcfit}, the estimated systematic error regions
are superposed on the data and best-fit model.

Using the pulse-profile model shown in Figure \ref{fig:maxilcfit},
we can calculate the pulse-phase-average X-ray flux for every 3-day
interval. In Figure \ref{fig:maxilcave}, the resulting 2--20\,keV
pulse-phase-average fluxes are plotted along with the raw GSC-scan data and the 2--20\,keV fluxes measured by NuSTAR. 
The time variation of the pulse-phase-average flux
is rather smooth.
In Figure \ref{fig:fgbmhist}, the pulse-phase-average MAXI/GSC 2--20\,keV flux is
compared with the Fermi/GBM 12--50\,keV pulsed flux. The good
correlation between the total average flux and the pulsed component
suggests that the pulsed fraction is largely constant throughout the
bright outburst period. The change in ratio between the MAXI flux and GBM pulsed flux throughout the outburst reflects the changing spectral hardness seen in Figure \ref{fig:maxilchr}.

\section{Estimating the Neutron Star magnetic field} \label{sec:bfield}

The timing and spectral analysis of \protag\ which we have presented yield several methods for the measurement of the magnetic field at the surface of the neutron star. First is the possible detection of a CRSF, which directly probes the magnetic field. The second method is to attribute the observed quasi-periodic oscillations to variability at the inner edge of the accretion disk. The third method is to investigate the relationship between the spin-up and luminosity throughout the observed outburst. Below, we elaborate on each of these methods and we discuss their respective advantages and disadvantages. 

\subsection{A tentative CRSF detection} \label{sec:crsf}

In Section \ref{sec:spectra}, we showed that the NuSTAR spectrum during OBS1 exhibited an absorption feature at 44\,keV that may be attributed to cyclotron resonance scattering. From \citet{Staubert2019}, we may write the relation between the energy of the fundamental cyclotron scattering feature in units of keV, $E_\mathrm{cyc}$, and the surface magnetic field strength in units of $10^{12}$\,G, $B_{12}$:

\begin{equation}
    E_\mathrm{cyc}=\frac{11.6}{1+z}\times B_{12}
\end{equation}

\noindent where $z$ is the gravitational redshift at the surface of the neutron star. For a canonical neutron star with mass $M=1.4M_\odot$ and radius $R=10$\,km, the value of the gravitational redshift is $1+z=1.31$. Thus, for a cyclotron energy at 44\,keV, the corresponding magnetic field strength is $B=5\times10^{12}$\,G.

It's important to note several caveats with respect to this spectral feature and therefore this magnetic field estimate. First, because the feature is in the the upper part of the NuSTAR bandpass, the appearance of this feature will inevitably be sensitive to the choice of continuum model. Second, the appearance of the feature in both of the CHU combinations which we chose was serendipitous. Upon investigating the spectra of other CHU combinations, we did not observe obvious absorption features at 44\,keV. Third, the line width, although poorly constrained, is narrow compared to typical cyclotron features at similar energies, which tend to have widths which lie between 4 and 10\,keV \cite{Staubert2019}. Given a line width of 2\,keV at 44\,keV we may estimate an electron velocity dispersion of $v/c\sim2/44 = 4.5\%$. This corresponds to a thermal energy of $\frac{1}{2}m_ev^2=0.5$\,keV. This is much lower than the typical  plasma temperatures inferred in other sources using models of thermal and bulk Comptonization \citep[e.g.][]{Iwakiri2019,Doroshenko2012}. If we consider, on the other hand, the possibility of proton scattering rather than electron scattering, the thermal broadening would require a temperature which is far too high at nearly a MeV ($\frac{m_p}{m_e}=1836$). Therefore a consideration of the width of the feature casts doubt on the feature, or at least on the hypothesis that it arises from scattering at the surface of the neutron star rather than in a region farther from the surface where the magnetic field is weaker and the plasma temperature lower.

Despite these considerations, we do not rule out the possibility of cyclotron scattering. Cyclotron features often show strong variability with pulse phase in their energies, widths, and depths. Given that the pulse period is of a similar timescale as NuSTAR's orbit, and therefore the timescale on which the operating CHU combination changes, variations in observed CRSF properties with CHU combination may not be unexpected. A thorough pulse-phase resolved analysis of the spectral variability during future outbursts may be necessary to produce a complete picture of the tentative CRSF which we report here. Applying such an analysis to OBS1 would require considerable care due to the idiosyncrasies of Mode 6 data, and we therefore consider it outside of the scope of this paper.

\subsection{QPO produced at the magnetospheric radius}\label{sec:qpobfield}

Millihertz QPOs, sometimes with harmonics, have been observed among accreting neutron stars in both LMXB and HMXB systems \citep[e.g.][]{James2010,Sidoli2016,Fei2021}. In neutron star LMXBs, which tend to host rapidly rotating accretors, low-frequency QPOs are sometimes attributed to Lense-Thirring precession \citep{Stella1998}. It is difficult to explain the QPO we have observed using this model, because the predicted Lense-Thirring frequency given a spin period of $\sim1100$\,s is far too low, on the order of $10^{-5}$\,Hz. Instead, in the case of disk accretion onto slowly-rotating, magnetized neutron stars in HMXB systems, this variability is often associated with the Keplerian rotation frequency at the inner edge of the truncated accretion disk. Kelvin-Helmholtz instabilities in the inner regions of the accretion disk may lead to modulation of the accretion rate onto the surface of the neutron star at the beat frequency of the Keplerian rotation frequency and the neutron star spin frequency \citep{Lamb1985}. Alternatively, inhomogeneities at the inner disk may lead to varying obscuration at the Keplerian rotation frequency \citep{vdKlis1987}.

We may therefore estimate the neutron star magnetic field if we assume that the low-frequency QPO we detect at $\nu=89$\,mHz is produced at or near the inner disk radius and that the disk is truncated at the magnetospheric radius. Due to the very small neutron star spin frequency compared to the QPO frequency, for either of the scenarios described above we may assume that the QPO frequency is equal to the Keplerian frequency at the inner disk:

\begin{equation}
    \nu_\mathrm{QPO} = \frac{1}{2\pi}\left(\frac{GM}{r^3_\mathrm{in}}\right)^{1/2}
\end{equation}

\noindent where $M$ is the mass of the neutron star and $r_\mathrm{in}$ is the inner disk radius. Furthermore, the magnetospheric radius, $r_m$, at which magnetic pressure balances the ram pressure of the accreting material, thereby disrupting the disk, is given by \citep{GhoshLamb1979II,GhoshLamb1979III}

\begin{equation}
    r_m=0.52\left(\frac{\mu^4}{2GM\dot{M}^2}\right)^{1/7}
\end{equation}

\noindent where $\dot{M}$ is the accretion rate and $\mu$ is the magnetic moment of the accreting neutron star. Lastly, the accretion rate can be related to the bolometric luminosity as measured at infinity by $L_\mathrm{bol} = \dot{M}(GM/R)(1+z)^{-1}$ if the majority of the luminosity originates from near the surface of the neutron star, which we will assume to be true given the very high pulse fraction. Combining all of these relations, we arrive at the following formula for the magnetic moment, $\mu$, in units of $10^{30}\,\mathrm{G\,cm^{3}}$:

\begin{equation}
\begin{aligned}
    \mu_{30} = 10.4 \times (1+z)^{1/2}\left(\frac{\nu_{\mathrm{QPO}}}{100\,\mathrm{mHz}}\right)^{-7/6} \\
    \times \left(\frac{L_\mathrm{bol}}{10^{37}\,\mathrm{erg\,s^{-1}}}\right)^{1/2}
    \left(\frac{R}{10\,\mathrm{km}}\right)^{1/2}\left(\frac{M}{M_\odot}\right)^{1/3}
\end{aligned}
\end{equation}

\noindent Assuming canonical values of $R=10$\,km and $M=1.4\,M_\odot$, and plugging in the values of $\nu_\mathrm{QPO}$ and $L_\mathrm{bol}$ that we measured during OBS1, we arrive at a magnetic moment of $\mu_{30}=11.4$, which corresponds to a surface magnetic field of $B=2.3\times10^{13}$\,G. Furthermore, this model would imply disk truncation at a radius of around $8000$\,km and, according to the equations of accretion disk structure introduced by \citet{ShakuraSunyaev}, an inner disk surface temperature of $kT\approx0.2$\,eV. These disk properties are consistent with the absence of a visible thermal disk component in the NuSTAR spectra.

\subsection{Luminosity - spin-up relation} \label{sec:ghoshandlamb}

\begin{figure}
\begin{center}
\includegraphics[width=0.49\textwidth]{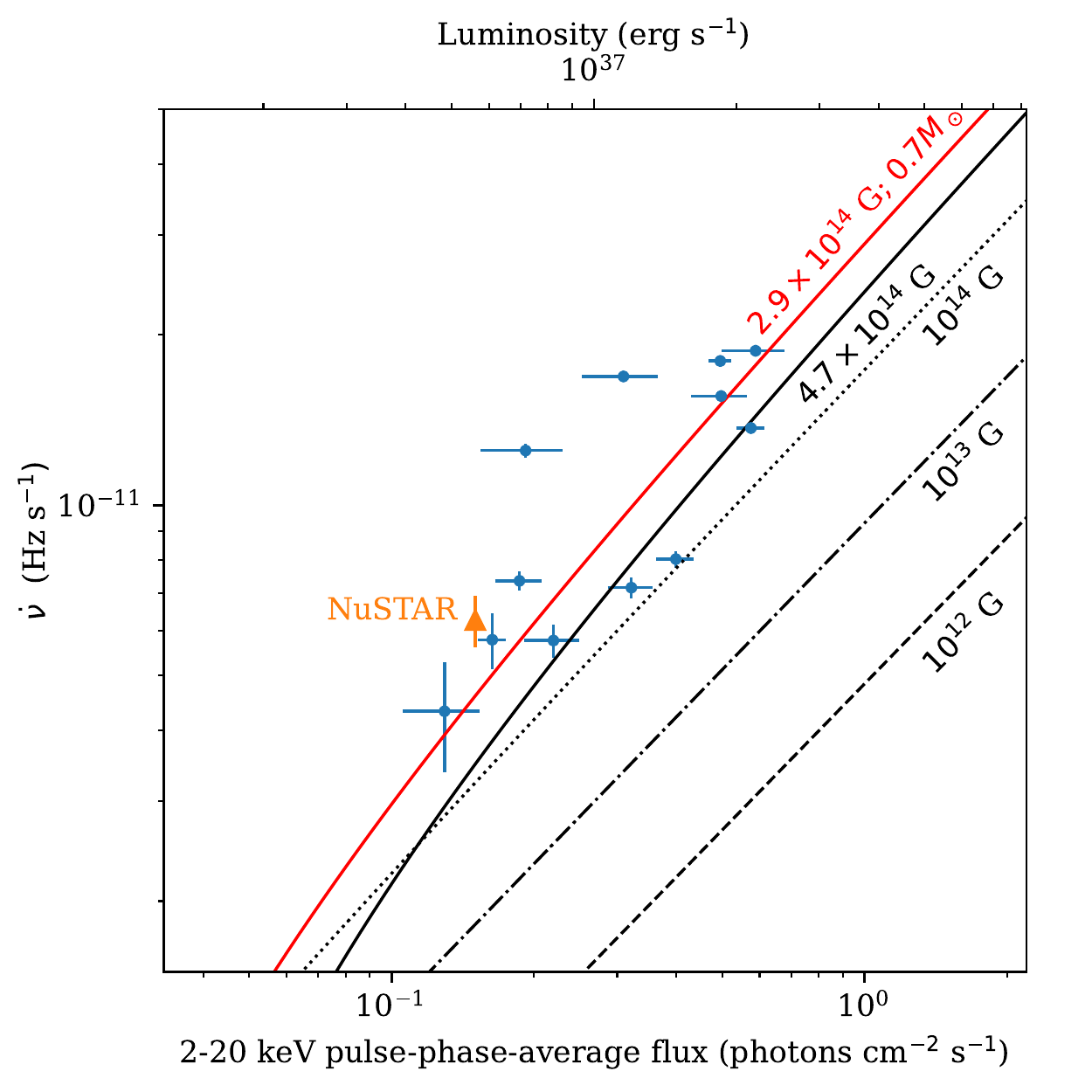}
\caption{
  Observed relation between the 2--20 keV pulse-phase-average photon flux 
  and pulse-frequency derivative $\dot{\nu}$, obtained from the MAXI, NuSTAR, Fermi/GBM 
  data analysis. 
  The horizontal scale at the top represents the expected source luminosity ($0.1-100$\,keV)
  given a source distance of 3.6\,kpc and the best-fit NuSTAR spectral model during OBS1.
  Theoretical predictions by the \citet{GhoshLamb1979III} model of accretion spin-up with 
  neutron star mass $1.4M_\sun$, radius 10\,km, spin period 1130\,s,
  and surface magnetic field $10^{12}$, $10^{13}$, $10^{14}$\,G,
  are drawn in dashed, dot-dashed, dotted lines, respectively. The line of best-fit to the data assuming a canonical neutron star mass of $M=1.4M_\odot$ is plotted as a solid black line and corresponds to a magnetic field of $4.7\times 10^{14}$\,G. If we assume an exceptionally low neutron star mass of $M=0.7M_\odot$, then higher spin-up rates are allowed by the model, and we retrieve a magnetic field strength of $2.9\times 10^{14}$\,G. This fit is plotted as a solid red line.
}
\label{fig:fdotlx}
\end{center}
\end{figure}




The third method by which we may estimate the magnetic field is via measurement of the torque exerted on the neutron star by the accreted matter. First of all, we must demonstrate that the observed variation in the spin period is not significantly impacted by Doppler shifting induced by the orbital motion of the accreting neutron star. The orbital periods of BeXRB pulsars typically lie in the range $P_\mathrm{orb}=20$--$200$\,days \citep[e.g.][]{2015A&ARv..23....2W}.
Assuming a O9.5-B0V companion mass of $18\,M_\odot$ \citep{Pecaut2013} Kepler's third law therefore yields an upper limit on the orbital velocity, 
$v_\mathrm{orb}/c\lesssim 7\times 10^{-4}$.
Additional considerations suggest an even lower limit. First, the empirical correlation between the spin and orbital periods in BeXRB pulsars
\citep{2017ApJ...839..119Y}
suggests that the orbital period of \protag\ should be $P_\mathrm{orb}\gtrsim 200$\,days.
Second, for $v_\mathrm{orb}/c\ll 1$, the ratio of the Doppler frequency shift
$\Delta\nu_\mathrm{orb}$ to the source frequency $\nu$ is approximated
only by the line-of-sight component of $v_\mathrm{orb}$, $v_\mathrm{orb}\cos{\theta}$.
The observed pulse-frequency change from 0.884 mHz
to 0.924 mHz during the observed outburst (Figure \ref{fig:fgbmhist})
corresponds to $\Delta\nu/\nu=4\times
10^{-2}$, which is larger by two orders of magnitude than the possible
orbital modulation estimated above.  
Therefore, we conclude that the observed
pulse-frequency change approximately represents that of the pulsar
spin-frequency $\nu_\mathrm{s}$. We note, however, that the ``instantaneous" frequency derivative (i.e. measured over short intervals) may be significantly affected by the orbital motion. For the binary parameters we considered above, the orbital contribution to the spin-up could reach $\dot{\nu}_\mathrm{orb}\approx2\times10^{-12}\,\mathrm{Hz\,s^{-1}}$ --- more than $10\%$ of the frequency derivative we measured during OBS1 --- for circular orbits, or even greater for eccentric orbits. Therefore, in the following analysis of accretion-induced spin-up we only consider secular frequency derivatives (i.e. measured across several days), in order to minimize the contribution of Doppler shifting due to the orbital motion of the pulsar.

Figure \ref{fig:fdotlx} shows the observed relation between the 2--20
keV pulse-phase-average flux obtained from the MAXI/GSC data and the
pulse-frequency derivative $\dot{\nu}$, i.e. the pulsar spin-up rate
from the Fermi/GBM pulsar data.
The $\dot{\nu}$ value for each data point is calculated with the
model function of the sequential pulse-frequency change (Figure \ref{fig:fgbmhist})
used in the pulse-folding analysis. 
The 2--20 keV flux during OBS1 estimated from the best-fit spectral model (Table
\ref{tab:spectral_params}) is also shown.
The figure clearly reveals a positive correlation between the spin-up rate and the flux.

We calculated the bolometric luminosity $L_\mathrm{bol}$ for the
observed 2--20\,keV flux, assuming a source distance of 3.6\,kpc, 
that the X-ray spectrum is consistent with the model in the
OBS1 (Table \ref{tab:spectral_params}), and
that emission is isotropic in pulse-phase average. In Figure \ref{fig:fdotlx}, the scale of
$L_\mathrm{bol}$ corresponding to the 2--20\,keV flux is shown on the top axis.
The observed 2--20\,keV flux range of $0.1\sim 0.7$ photons cm$^{-2}$\,s$^{-1}$ corresponds to $L_\mathrm{bol} = (0.4\sim2) \times 10^{37}$
erg\,s$^{-1}$, which is around $\sim 1/10$ of the Eddington
luminosity of the nominal $1.4\,M_\sun$ neutron star.

Then, we compared the measured $\dot{\nu}$-$L_\mathrm{bol}$
relation with theoretical models of pulsar spin-up by accreting matter coupled to the neutron star
magnetic field.
Several different models have been proposed 
\citep[e.g.][]{GhoshLamb1979III,1995MNRAS.275..244L,2007ApJ...671.1990K}.
Among these models, the predicted $\dot{\nu}$-$L_\mathrm{bol}$ relations are slightly different by a factor of $\sim 2$ \citep[e.g.][]{2009A&A...493..809B}.
We here employed a model of disk-magnetosphere interaction 
proposed by \citet[][GL hereafter]{GhoshLamb1979III}, which predicts a proportionality of $\dot{\nu} \propto L_\mathrm{bol}^{6/7}$,
because the GL model has been compared against observational data
and its predictive power has been demonstrated \citep[e.g.][]{2017PASJ...69..100S}.
%
%
In Figure \ref{fig:fdotlx},
the expected relations from the GL model for a neutron star with spin period 1130\,s and canonical mass ($1.4M_\odot$), radius (10\,km), and moment of inertia ($10^{45}$\,g\,cm$^{2}$),
are shown using dashed, dot-dashed, and dotted lines, corresponding to surface magnetic fields of  $B=10^{12}, 10^{13}, 10^{14}$\,G, respectively. 

The best fit to the data, assuming a mass of $1.4\,M_\odot$ and radius of 10\,km, is shown as a solid black curve in Figure \ref{fig:fdotlx} and corresponds to a magnetic field strength of $4.7\times10^{14}$\,G. We find that the GL model cannot achieve a good fit given the assumed parameters. The cause of this is that the observed spin-up rates are near the maximum that can be produced by the GL model for the accretion rates (i.e. luminosities) which we have inferred from the observed flux. For a constant mass accretion rate corresponding to $L\approx10^{37}\,\mathrm{erg\,s^{-1}}$, the $\dot{\nu}$-$B_{\mathrm{s}}$ relation predicted by the GL model peaks at $\dot{\nu}\approx5\times10^{-12}$\,Hz for a magnetic field of $B_{\mathrm{s}}\approx5\times10^{14}$\,G. 

This mismatch between the data and model can be ameliorated by a change of assumptions. For example, if we assume a lower neutron star mass, then the same accretion rates will result in more rapid spin-up. We show the best fit to the data assuming a neutron star mass of $0.7\,M_\odot$ as a solid red line in Figure \ref{fig:fdotlx}. This minimal value of the mass results in a best-fit value of $2.9\times10^{14}$\,G. While the neutron star may indeed have a mass lower than $1.4\,M_\odot$ (though perhaps not as low as $0.7\,M_\odot$) it is also likely that there is significant statistical uncertainty in our estimation of the intrinsic luminosity of the source. In calculating the luminosity from the observed flux, we have assumed isotropic emission, while the highly pulsed flux demonstrates that the emission is highly anisotropic. Additionally, we translated flux into luminosity assuming the spectral shape and flux observed during NuSTAR OBS1. Spectral evolution throughout the outburst will affect the positions of the data points in Figure \ref{fig:fdotlx} along the x-axis and may have resulted in the spread in luminosities for similar spin-up rates. Nonetheless, the proportional relationship between the fluxes and spin-up rates we have observed corresponds to a magnetic field strength on the order of $B_\mathrm{s}\sim10^{14}$\,G.



\section{Summary and Conclusions} \label{sec:conclusions}

We have presented a comprehensive view of the 2022 outburst of the BeXRB, \protag. The outburst was detected by MAXI, and due to the angular proximity of the source to the Sun, rapid follow-up was only possible with NuSTAR. We performed a spectral and timing analysis of the NuSTAR data which revealed pulsations with a period of around 1100\,s and a rapid secular spin-up over the course of the $\sim50$\,day outburst. Given the length of the outburst and the peak luminosity, around $L_\mathrm{bol}\approx2\times10^{37}\,\mathrm{erg\,s^{-1}}$, this outburst is consistent with being a type II outburst. 

The source showed strong pulsations with a pulse fraction of 65--85\% throughout the outburst. Non-sinusiodal pulse profiles indicate a complex magnetic field structure which warrants further inspection in future work. While the MAXI light-curve apparently exhibited dramatic flaring behavior, by combining NuSTAR pulsation measurements with those of Fermi/GBM, we were able to reconstruct the pulse-averaged light-curve in order to show that the outburst followed a much smoother evolution and that the days-timescale variability seen in the MAXI light-curve was not intrinsic to the source. 

Additionally, we have estimated the surface magnetic field $B_\mathrm{s}$ of the neutron star
using three independent methods:
\begin{enumerate}
\item[(I)] $5\times 10^{12}$\,G from the possible CRSF at 44\,keV in the NuSTAR X-ray spectrum,
\item[(II)] $2.3\times 10^{13}$\,G from the 89\,mHz QPO in the NuSTAR X-ray variability,
\item[(III)] $\sim10^{14}$\,G from the $\dot{\nu}$-$L_\mathrm{bol}$ relation during the present outburst observed by the MAXI/GSC and Fermi/GBM.
\end{enumerate}
While (I) agrees with the typical values of known BeXRB pulsars, $(1-8)\times 10^{12}$ G
\citep[e.g.][]{NPEX}, the other two are significantly higher, and in fact cyclotron resonance at a magnetic field strength greater than $10^{13}$\,G would produce a scattering features at energies $\gtrsim 90$\,keV, beyond the NuSTAR bandpass.

Measurements of neutron star magnetic fields using CRSFs ostensibly depend only on fundamental physical processes. Therefore, there is no need to consider systematic error associated with the method, and the statistical error is low, on the order of only a few percent. On the other hand, the statistical significance of the observed feature, $<3\sigma$ may not be considered sufficient evidence, and other considerations, such as the unusually narrow width of the feature, cast further doubt on the significance of the feature and on its hypothesized origin from cyclotron scattering at the neutron star surface.

The second method assumes that the frequency of QPOs represent the Kepler orbit at the radius at which the accretion disk is truncated by the dipole magnetic fields of the neutron star \citep[e.g.][]{1996ApJ...459..288F}.
The estimated value of the magnetic field therefore depends on the assumed theoretical model about the disk-magnetosphere interaction
\citep[e.g.][]{GhoshLamb1979III}
and also includes uncertain factors like the bolometric correction factor, emission anisotropy, and source distance. 
At this time, QPOs with frequencies in the range $\sim$10--100\,mHz have been observed in a handful of BeXRB pulsars \citep{James2010},
and a few of them are considered to be consistent with the present scenario 
\citep[e.g.][]{1996ApJ...459..288F,2011MNRAS.417..348D,2021MNRAS.508.5578R}. 
Among this sample, the estimated surface magnetic field strength is consistent with 
those derived from the CRSF measurements within a factor of a few ($\sim 2$).
If this discrepancy accurately reflects the systematic error inherent to this method, 
our obtained value of $2.3\times 10^{13}$\,G allows 
for a range of values from $1.2\times 10^{13}$\,G to $4.6\times 10^{13}$\,G. 
%
Hence, if the observed QPO really represents the Keplerian orbital frequency at the inner-disk radius,
$B_\mathrm{s}$ is estimated to be $\gtrsim 10^{13}$\,G including the systematic uncertainty.

The third method utilizes the predicted relation between bolometric luminosity and spin-up rate.
The $\dot{\nu}$-$L_\mathrm{bol}$ relation we observe shows a clear correlation,
therefore agreeing with theoretical models that
the pulsar is spun up by accreting matter through the neutron star
magnetic field.
Like the QPO method, this estimate also depends on theoretical predictions of disk-magnetosphere interaction. We discussed various sources of systematic error in Section \ref{sec:ghoshandlamb}, and indeed, using the MAXI/GSC and Fermi/GBM data for 12 BeXRB pulsars,
the factor of proportionality between $\dot{\nu}$ and $L_\mathrm{bol}^{6/7}$ in the GL model 
has been shown to disagree with observations 
by up to a factor of $\sim 3$ \citep{2017PASJ...69..100S}.
Given that the coefficient of proportionality between $\dot{\nu}$ and $L_\mathrm{bol}^{6/7}$ 
itself goes as $\mu_{30}^{2/7}$, the estimated $B_\mathrm{s}$ ($\propto \mu_{30}$) may include a systematic error of factor $\sim 3^{7/2}\simeq 50$ in the worst case.
Even if we include this error factor in our estimate, the best-fit value of $B_\mathrm{s}=4.7\times 10^{14}$\,G suggests that the true $B_\mathrm{s}$ is at least $10^{13}$ G.

%
With a pulse period of $P_\mathrm{s}\approx1100$\,s, \protag\  is 
among the slowest-spinning BeXRB pulsars 
in our Galaxy \citep[e.g.][]{2015A&ARv..23....2W}.
If the rate of rotation is close to the spin equilibrium 
and the Keplerian orbital period at the inner disk edge is close to the pulsar spin period, 
the long spin period suggests a strong dipole magnetic field:
\begin{equation}
\begin{split}
B_\mathrm{s} & = 1.6\times 10^{12} 
\left(\frac{P_\mathrm{s}}{1\,\mathrm{s}}\right)^{7/6}  
\left(\frac{L_\mathrm{bol}}{10^{37}\,\mathrm{erg\, s}}\right)^{1/2}  \\
& \quad \times \left(\frac{R_\mathrm{NS}}{10\,\mathrm{km}}\right)^{-5/2} 
\left(\frac{M_\mathrm{NS}}{1.4 M_\sun}\right)^{1/3} \quad \mathrm{G}
\end{split}
\end{equation}
In fact, X Persei --- a similarly slow-spinning BeXRB pulsar with a pulse period of 837\,s ---
has also been suggested to have a magnetar-like field strength, $B_\mathrm{s}\gtrsim 4\times 10^{13}$ 
\citep{2018PASJ...70...89Y}.
\protag\ may be another example of this kind of slow-spinning magnetar-like BeXRB pulsar. NuSTAR observations of future outbursts from this source, as well as thorough pulse-profile modeling and pulse-phase-resolved spectroscopic studies, may help to further elucidate the magnetic field of this slowly-rotating neutron star.

\begin{acknowledgments}
SNP acknowledges partial support from the National Aeronautics and Space Administration (NASA) via the NuSTAR General Observer Program grant number 80NSSC23K0399.
JvdE acknowledges a Warwick Astrophysics prize post-doctoral fellowship made possible thanks to a generous philanthropic donation and was supported by a Lee Hysan Junior Research Fellowship awarded by St Hilda's College, Oxford during part of this work.
MS acknowledges partial support from the Chinese Academy of Sciences (CAS) President's International Fellowship Initiative (PIFI) (grant No. 2020FSM004).
Part of this work was financially supported by Grants-in-Aid for Scientific Research 21K03620 (HN) and 19K14762 (MS) from the Ministry of Education, Culture, Sports, Science, and Technology (MEXT) of Japan.

This research has made use of the NuSTAR Data Analysis Software (NuSTARDAS) jointly developed by the ASI Science Data Center (SSDC, Italy) and the California Institute of Technology (USA).
We thank the staff at the South African Radio Astronomy Observatory (SARAO) for scheduling the MeerKAT observations. The MeerKAT telescope is operated by the South African Radio Astronomy Observatory, which is a facility of the National Research Foundation, an agency of the Department of Science and Innovation. This work was carried out in part using facilities and data processing pipelines developed at the Inter University Institute for Data Intensive Astronomy (IDIA). IDIA is a partnership of the Universities of Cape Town, of the Western Cape and of Pretoria.

We thank the anonymous referee for their comments and suggestions which improved this work.

\end{acknowledgments}

%

\vspace{5mm}
\facilities{MAXI, NuSTAR, Fermi, MeerKAT}


\software{astropy \citep{astropy2013,astropy2018}, Xspec \citep{Xspec}, Stingray \citep{stingray_doi,Huppenkothen2019a,Huppenkothen2019b}, SciPy \citep{Scipy}, DS9 \citep{ds9}}

\bibliography{main}{}
\bibliographystyle{aasjournal}



\end{document}